\documentclass[a4paper,fleqn,sort&compress]{cas-sc}

\usepackage[utf8]{inputenc}
\usepackage[english]{babel}
\usepackage[square,numbers]{natbib}
\usepackage{placeins}
\usepackage{amsmath}
\usepackage{graphicx}
\usepackage{tcolorbox}
\usepackage{relsize}
\usepackage{comment}
\usepackage{tabu,multirow}
\usepackage{epsfig}
\usepackage{array}
\usepackage{amssymb}
\usepackage{latexsym}
\usepackage{xcolor}
\usepackage{subfigure}
\usepackage{soul}
\usepackage[export]{adjustbox}
\usepackage{setspace}
\usepackage{caption}
\usepackage{float}
\newcolumntype{P}[1]{>{\centering\arraybackslash}p{#1}}
\newcommand{\beq}{\begin{equation}}
\newcommand{\eeq}{\end{equation}}
\newcommand{\beqs}{\begin{eqnarray}}
\newcommand{\eeqs}{\end{eqnarray}}
\newcommand{\itbf}[1]{\textit{\textbf{#1}}}

\newcommand{\boldface}[1]{\boldsymbol{#1}}  

\newcommand{\bfd}{\boldface{d}}
\newcommand{\bfe}{\boldface{e}}

\newcommand{\bfp}{\boldface{p}}
\newcommand{\bfq}{\boldface{q}}

\newcommand{\bfu}{\boldface{u}}

\newcommand{\bfz}{\boldface{z}}

\newcommand{\bfF}{\boldface{F}}

\newcommand{\bfI}{\boldface{I}}

\newcommand{\bfepsilon}{\boldsymbol{\varepsilon}}

\newcommand{\bfnull}{\boldsymbol{0}}

\newcommand{\bfSigma}{\boldsymbol{\Sigma}}

\newcommand{\bfOmega}{\boldsymbol{\Omega}}

\newcommand{\calH}{\mathcal{H}}

\newcommand{\calZ}{\mathcal{Z}}
\newcommand{\T}{^{\mathrm{T}}} 

\newlength{\boxwidth}
\def\dd{\;\!\mathrm{d}}

\def\btheorem{\begin{theorem}}
\def\etheorem{\end{theorem}}
\def\blemma{\begin{lemma}}
\def\elemma{\end{lemma}}
\def\bproposition{\begin{proposition}}
\def\eproposition{\end{proposition}}
\def\bcorollary{\begin{corollary}}
\def\ecorollary{\end{corollary}}
\def\bdefinition{\begin{definition}}
\def\edefinition{\end{definition}}
\def\bexample{\begin{example}}
\def\eexample{\end{example}}
\def\bremark{\begin{remark}}
\def\eremark{\end{remark}}

\def\id{{\bfI}}

\DeclareMathOperator{\tr}{tr}

\newcommand{\be}{\begin{equation}}
\newcommand{\ee}{\end{equation}}

\newcommand{\bem}{\begin{multline}}
\newcommand{\eem}{\end{multline}}
\newcommand{\ba}{\begin{align}}
\newcommand{\ea}{\end{align}}

\newcommand{\RA}{\qquad\Rightarrow\qquad}

\begin{document}
\let\WriteBookmarks\relax
\shorttitle{Finite-temperature surface elasticity of crystalline solids}
\def\floatpagepagefraction{1}
\def\textpagefraction{.001}
\shortauthors{Saxena et~al.}

\title [mode = title]{Finite-temperature surface elasticity of crystalline solids}

\author[1]{Shashank Saxena}[orcid=0000-0002-5242-9103]
\author[1]{Miguel Spinola}[orcid=0000-0002-5180-6149]
\author[2]{Prateek Gupta}[orcid=0000-0003-3666-0257]
\ead{prgupta@am.iitd.ac.in}
\author[1]{Dennis M. Kochmann}[orcid=0000-0002-9112-6615]
\cormark[1]
\ead{dmk@ethz.ch}
\ead[URL]{mm.ethz.ch}
\address[1]{Mechanics \& Materials Lab, Department of Mechanical and Process Engineering, ETH Z\"urich, 8092 Z\"urich, Switzerland}
\address[2]{Department of Applied Mechanics,  Indian Institute of Technology Delhi, 110016, New Delhi, India}

\cortext[cor1]{Corresponding author}










\begin{abstract}
Surface energies and surface elasticity largely affect the mechanical response of nanostructures as well as the physical phenomenon associated with surfaces such as evaporation and adsorption. Studying surface energies at finite temperatures is therefore of immense interest for nanoscale applications. However, calculating surface energies and derived quantities from atomistic ensembles is usually limited to zero temperature or involve cumbersome thermodynamic integration techniques at finite temperature. Here, we illustrate a technique to identify the energy and elastic properties of surfaces of solids at non-zero temperature based on a Gaussian phase packets (
GPP) approach (which in the isothermal limit coincides with a maximum-entropy formulation). Using this setup, we investigate the effect of temperature on the surface properties of different crystal faces for six pure metals -- copper, nickel, alumimum, iron, tungsten and vanadium -- thus covering both FCC and BCC lattice structures. While the obtained surface energies and stresses usually show a decreasing trend with increasing temperature, the elastic constants do not show such a consistent trend across the different materials and are quite sensitive to temperature changes. Validation is performed by comparing the obtained surface energy densities of selected BCC and FCC materials to those calculated via molecular dynamics.
\end{abstract}




\maketitle

\section{Introduction}

Surfaces of solids are known to have an energy in excess to that of the bulk, which can be attributed to a lower coordination number of the surface atoms. The effect of this surface energy is in proportion to the surface-to-volume ratio of a sample. Hence, although negligible in large-scale structures, surfaces may have a pronounced effect, when it comes to understanding the mechanical properties of nano-scale structures. Experimental \cite{gao2000nanomechanics,yang2002energy} and numerical \cite{duan2005size,liang2005size,amelangkochmann2015} findings have shown that the stiffness of mechanical structures tends to deviate from its known large-scale values as one or more sample dimensions approach the material length scale, which is the lattice parameter for crystalline solids. For example, the elastic response of plates with thickness approaching a few atomic planes is drastically different than what that of thicker specimens. As shown by \citet{zhou2004surfaces}, such nanostructures can be either softer or stiffer compared to the bulk.
Studying surface energies is of scientific interest for a wide range of physical phenomena associated with surfaces, going beyond mechanics. Size effects have been demonstrated to affect physical properties such as the melting point \cite{wang2003melting,shim2002thermal} and the onset temperature of evaporation \cite{nanda2003higher} for nanoparticles. Surface energy also plays an important role in the phase transformation of gold nanowires \cite{gall2005tetragonal}. Moreover, the strong growth of nanotechnology and the use of nanoelectromechanical systems (NEMS) provide for a growing branch of engineering that requires a careful understanding of surface energetics. 

In a continuum setting, the effect of surface energies can be captured by considering the presence of an infinitely-thin deformable surface layer covering the bulk material \cite{gibbs1906scientific}, which is endowed with surface stresses and associated surface elastic constants. Important works in the development of continuum surface theories include those of \citet{shuttleworth1950surface} and \citet{gurtin1975continuum}, which allowed for the calculation of surface stress and elasticity tensors from atomistics by calculating derivatives of the computed surface energies with respect to the applied strain. For amorphous materials, attempts to study surface stresses have been mainly experimental \cite{xu2018surface,schulman2018surface} because of the difficulties in their numerical simulations owing to their microstructure without long-range order. These works utilize the contact angle measurements of glycerol drops on solid surfaces to find the surface stress. For soft solids, surface stresses may dominate the elastic energy, leading to physical phenomena associated with \textit{elastocapillarity} \cite{style2017elastocapillarity}. These effects become prominent below a critical length scale determined by the ratio of the surface stress to the elastic modulus. Common examples include surface stress-induced rounding of sharp surface profiles \cite{jagota2012surface,hui2020surface}, reduced indentation depths and contact radii compared to those determined by contact theories that neglect surface effects \cite{style2013surface}, and higher fracture resistance due to crack tip blunting \cite{kundu2009cavitation, liu2014energy}. Some of these experimentally measurable phenomena are in turn used to quantify surface stresses in soft solids. For crystalline solids, experimental works \citep{wasserman1972determination,mays1968surface} have related the free surface stresses to a lattice parameter contraction measured by electron diffraction techniques. In addition, surface energy has also been related to the heat of formation of metals \citep{de1988cohesion}. Such experimental studies have been complemented by an abundance of computational investigations. These include, among others, \textit{ab initio} calculations \cite{wang2019effects,schonecker2015thermal}, Monte-Carlo (MC) \cite{frolov2009temperature}, Molecular Statics (MS) \cite{sheng2011highly,shenoy2005atomistic}, and Molecular Dynamics (MD) studies \cite{xing2020temperature,grochola2002new,grochola2002simulation,grochola2004exact}. \citet{miller2000size} showed that the relative difference between the mechanical properties of nanostructures and the corresponding bulk properties follows an inverse relation with the characteristic length scale of the nanostructure. They verified this observation with atomistic data for extension and bending of nanoplates and nanorods. \citet{shenoy2005atomistic} demonstrated the effect of surface relaxation on computationally obtained surface elastic constants and proposed an accurate (albeit expensive) procedure for finding surface energies of FCC metals under imposed strains. Recently, \citet{sievers2020computational} reported a computational homogenisation technique to find the surface elastic constants by establishing an energy equivalence between atomistic and continuum representative volume elements (RVEs) at zero temperature.

It is important to note that the majority of computational studies involved MS simulations at zero temperature (0~K). While this is computationally convenient, it is far from physical reality and does not easily admit extensions to non-zero temperature. A finite-temperature estimation of the surface properties from MD has been accomplished \cite{xing2020temperature,grochola2002simulation,freitas2017step} using the procedure of thermodynamic integration as illustrated by \citet{freitas2016nonequilibrium}, originally introduced by \citet{frenkel1984new}. Although beneficial, this approach is unfortunately rather cumbersome in practice, as will be discussed later. Further MD studies on finding the size-dependent elastic constants of nanowires and nanobeams at finite temperature exist (see, e.g., \cite{gall2004strength,wu2006molecular,zhang2009influence}), but these do not quantify the specific effect of the surface alone and hence to not lend to a transition to the continuum theories of solids with surface energies. 

Our objective is an accurate and computationally inexpensive methodology for obtaining surface properties of crystalline solids at finite temperature. We follow the Gaussian Phase Packets (GPP) formalism recently introduced by \citet{gupta2021nonequilibrium}, which captures the effect of thermal vibrations via a statistical-mechanics approach by considering a Gaussian distribution of atomic positions and momenta and finding the quasistatic equations of equilibrium for the distribution parameters at finite temperature. For isothermal conditions, this approach was shown to be equivalent to the finite-temperature maximum-entropy framework of \citet{KulkarniEtAl2008} and \citet{VenturiniEtAl2014} for long-term atomistic simulations. We validate our results for representative cases by comparison to MD data obtained using the \textit{Large-scale Atomic/Molecular Massively Parallel Simulator} (LAMMPS) \cite{LAMMPS} and thermodynamic integration \cite{freitas2016nonequilibrium}.

The remainder of this contribution is structured as follows. Section~\ref{GPP review} briefly summarizes the statistical GPP approach, highlighting the key points relevant for this work. Section~\ref{methodology} describes the methodology for obtaining surface energies, surface stresses, and surface elastic constants along with simulation details. There, we also provide a brief summary of nonequilibrium thermodynamic integration \cite{freitas2016nonequilibrium}, as used in LAMMPS. Next, Section~\ref{results} presents all numerical results in two parts. We first demonstrate the validity of our methodology to obtain geometrically converged surface free energies for various FCC and BCC metals and surface orientations at finite temperatures with comparison to LAMMPS data  (Subsection~\ref{Thickness convergence}). Second, we present surface stresses and surface elastic constants for all materials studied (Subsection~\ref{surface elastic constants results}). Finally, Section \ref{conclusion} concludes this study and discusses future directions.

\section{Gaussian Phase Packet formulation: a review} 
\label{GPP review}

Let us briefly review the recently introduced Gaussian Phase Packet (GPP) formulation by \citet{gupta2021nonequilibrium} to the extent necessary for subsequent discussions. While traditional MD is a robust and well established method to study nanoscale systems, it often suffers from long statistical convergence times at finite temperature. The GPP approach derives inspiration from statistical mechanics, where the thermodynamic quantities of interest (such as surface or bulk energies) can be linked to phase averages depending on the positions $\bfq=\{\bfq_i(t) : i=1,\ldots,N\}$ and momenta  $\bfp=\{\bfp_i(t) : i=1,\ldots,N\}$ of all $N$ atoms in an ensemble. Determining those parameters from the statistically averaged equations of motion is the main objective of the GPP approach. Following \citet{ma1993approximate}, the GPP formulation posits a multivariate Gaussian form of the probability distribution function, i.e.,
\begin{equation}
    f(\bfz ,t) = \frac{1}{\calZ(t)} \exp\left[ -\frac{1}{2}(\bfz  - \bar{\bfz }(t))^T \boldsymbol{\Sigma}^{-1}(t) (\bfz  - \bar{\bfz }(t))\right],
\end{equation}
where $\bfz=\left(\bfp(t),\bfq(t)\right)  \in \mathbb{R}^{6N}$ is a condensed representation of the phase-space coordinate, $\bar{\bfz }(t)=
\langle\bfz\rangle=\int f(\bfz,t)\bfz\dd\bfz$ denotes the mean phase space coordinate, and $\bfSigma\in \mathbb{R}^{6N\times 6N}$ is the covariance matrix of interatomic positions and momenta. $\calZ(t)$ is the partition function, defined via $\int f(\bfz,t)\dd z = \langle 1\rangle = 1$, integrating over all of phase space.

Inserting the above probability distribution into Liouville's equation leads to a time evolution problem \cite{gupta2021nonequilibrium} and defines the equations of motion for the mean positions and momenta, $\bar{\bfz }$, as well as for the statistical information contained in $\boldsymbol{\Sigma}$. In the most general case, $\boldsymbol{\Sigma}$ is a fully populated matrix, whose off-diagonal terms play a crucial role in interatomic energy transfer. However, the resulting evolution equations to be solved for $\bar{\bfz }$ and $\boldsymbol{\Sigma}$ are numerically even stiffer than the traditional MD equations. Hence, interatomic independence (i.e., $\Sigma_{ij}=0$ for $i\neq j$) is assumed to solve for the equilibrium configuration of a system, while the interatomic correlations responsible for nonequilibrium irreversible thermal transport must be modelled separately. Since our focus here is on the equilibrium configurations of strained or unstrained surfaces, we limit ourselves to independent Gaussian phase packets, which implies
\begin{equation}
    f_i(\bfz _i,t) = \frac{1}{\calZ _i(t)} \exp\left[ -\frac{1}{2}(\bfz _i - \bar{\bfz _i}(t))^T \boldsymbol{\Sigma}_i^{-1}(t) (\bfz _i - \bar{\bfz _i}(t))\right] \qquad \text{such that} \qquad f(\bfz ,t) = \prod_{i=1}^N f_i(\bfz _i,t).
    \label{independent GPP}
\end{equation}

Further assuming a hyperspherical shape of the atomic distribution function $f _i$ in six dimensions leads to vanishing correlations of positions and momenta in different directions. The only non-zero terms thus remaining in the covariance matrix are (with $\tr(\cdot)$ denoting the trace of a matrix)
\begin{equation}
    \Omega_i = \frac{1}{3}\text{tr}\left( \boldsymbol{\Sigma}_i^{(\bfp ,\,\bfp )}\right), \qquad \Sigma_i = \frac{1}{3}\text{tr}\left( \boldsymbol{\Sigma}_i^{(\bfq ,\,\bfq )}\right), \qquad\mathrm{and}\qquad \beta_i = \frac{1}{3}\text{tr}\left( \boldsymbol{\Sigma}_i^{(\bfp ,\,\bfq )}\right),
\end{equation}
where we defined the covariance matrix of the type
\begin{equation}
 \bfSigma_{i} = \left(\begin{matrix}
                    \bfSigma^{(\bfp,\bfp)}_{i} &  \bfSigma^{(\bfp,\bfq)}_{i} \\
                     \bfSigma^{(\bfq,\bfp)}_{i} &  \bfSigma^{(\bfq,\bfq)}_{i}
                       \end{matrix}
\right),
\label{eq: lumped_covariance}
\end{equation}
consisting of diagonal block matrices.
Therefore, the set of parameters to be solved for each atom becomes $\left( \bar\bfp _i,\bar\bfq_i,\Omega_i,\Sigma_i,\beta_i\right) $. Their evolution over time is obtained from the phase-averaged equations of motion, which result from inserting the above into Liouville's equation \cite{gupta2021nonequilibrium}:
\be
\begin{split}
&\frac{\text{d} \bar{\bfq }_i}{\text{d}t} = \frac{\langle \bfp _i \rangle }{m_i},\qquad  \frac{\text{d} \bar{\bfp }_i}{\text{d}t} = \langle \itbf{F}_i  \rangle, \\
   & \frac{\text{d} \Omega_i}{\text{d}t} = \frac{ \langle \itbf{F}_i(\bfq ) \cdot (\bfp  - \bar{\bfp })  \rangle }{3}, \\
   &\frac{\text{d} \Sigma_i}{\text{d}t} = \frac{2\beta_i}{m_i},\\
   &\frac{\text{d} \beta_i}{\text{d}t} = \frac{\Omega_i}{m_i} + \frac{ \langle \itbf{F}_i(\bfq ) \cdot (\bfq  - \bar{\bfq })  \rangle }{3},
\end{split}
\ee
where $\bfF_i$ denotes the net force acting on atom $i$ having mass $m_i$.

In the quasistatic limit, mean momenta $\bar{\bfp }_i$ and mean thermal momenta $\beta_i$ vanish for every atom. Consequently, the evolution equation for $\Omega_i$ becomes an identity, and we are left with four scalar equations in the quasistatic limit:
\be\label{EOM quasistatic}
\begin{split}
&\langle \itbf{F}_i \rangle = \itbf{0}\qquad \mathrm{and} \\
&\frac{\Omega_i}{m_i} + \frac{ \langle \itbf{F}_i(\bfq ) \cdot (\bfq  - \bar{\bfq })  \rangle }{3} = 0,
\end{split}
\ee
whose solution is the set of average positions $\bar\bfq=\{ \bar\bfq _i: i=1,\ldots,N \}$ and position variances $\Sigma=\{\Sigma_i : i=1,\ldots, N\}$ for all atoms in equilibrium. Importantly, this admits decoupling the phase dynamics of thermal vibrations from the slow mean motion of atoms, which is essential towards our objective of studying equilibrium properties at finite temperature. Rather than resolving atomic motion at the femtosecond level, this approach tracks the effective atomic parameters $\left( \bar\bfp _i,\bar\bfq_i,\Omega_i,\Sigma_i,\beta_i\right) $ over time (at significantly larger time scales than required for $\left(\bfp _i,\bfq_i\right) $ in MD) and, in the quasistatic limit, reduces to a set of equilibrium equations to be solved for the aforementioned effective parameters.

Information about the momentum variances $\Omega=\{\Omega_i:i=1,\ldots,N\}$ is obtained from the type of thermodynamic process \textit{assumed} to bring the system to equilibrium. The entropy of the system is obtained using Boltzmann's expression,
\be
    S = -k_B \langle \ln f \rangle = \sum_{i=1}^N \Bigg[S_{0_i} - 3k_B \ln h + \frac{3k_B}{2} \ln(\Omega_i \Sigma_i) \Bigg],
\ee
where $S_{0_i} = 3 k_B [1+\text{ln}(2\pi)] - \text{ln}(N!)/N$, $h$ is Planck's constant, and $k_B$ Boltzmann's constant. Therefore, to simulate, e.g., an isentropic process, condition $\Omega_i \Sigma_i = \text{const.}$ for each atom $i$ complements Eqs.~\eqref{EOM quasistatic}. In this work, we will focus on isothermal simulations, for which the condition $\Omega_i = m k_B T$ holds for every atom $i$ in a system at a constant temperature~$T$.

The system of equations in \eqref{EOM quasistatic} can also be interpreted as stationarity conditions, aiming to find the minimizer of the Helmholtz free energy of the system, defined by 
\begin{equation}
    \mathcal{F}(\bar{\bfq },\Omega,\Sigma) = E (\bar{\bfq },\Omega,\Sigma) - \sum_{i=1}^N \frac{\Omega_i S_i}{k_B m_i},
    \label{Helmholtz free energy}
\end{equation}
where we introduced the internal energy of the system as the average total Hamiltonian
\begin{equation}
    E (\bar{\bfq },\bfOmega,\Sigma) =  
    \sum_{i=1}^N \frac{3\Omega_i}{2 m_i} + \langle V_i(\bfq ) \rangle.
\end{equation}
Here, $V(\bfq )$ is the total potential energy of the system, typically defined via interatomic potentials (as in this study). 
 

Eq.~\eqref{Helmholtz free energy} represents a convenient way to calculate the Helmholtz free energy $\mathcal{F}$ from our GPP-based simulations, which is beneficial since $\mathcal{F}$ is the relevant thermodynamic potential of interest for finite-temperature calculations with imposed strains (and considerably harder to extract from MD, as explained in Section~\ref{methodology}). For a detailed description of the relevant Legendre transforms of the internal energy corresponding to the natural control variables for the thermodynamic process, see, e.g., \citep{alberty2001use}. In the following, we refer to the Helmholtz free energy as `\textit{free energy}' for conciseness.

\section{Methodology} 
\label{methodology} 

In this section, we introduce the procedures used to isolate surface free energies and derived quantities (such as surface stresses and elastic constants) from atomistic simulations, using both the GPP approach and MD with thermodynamic integration. We first outline the general procedure and simulation domain and then discuss simulation details for each of the two approaches.

\subsection{Slab configuration and surface energy definitions}
\label{sec:Configuration}

\begin{figure}
\centering
\epsfig{file = 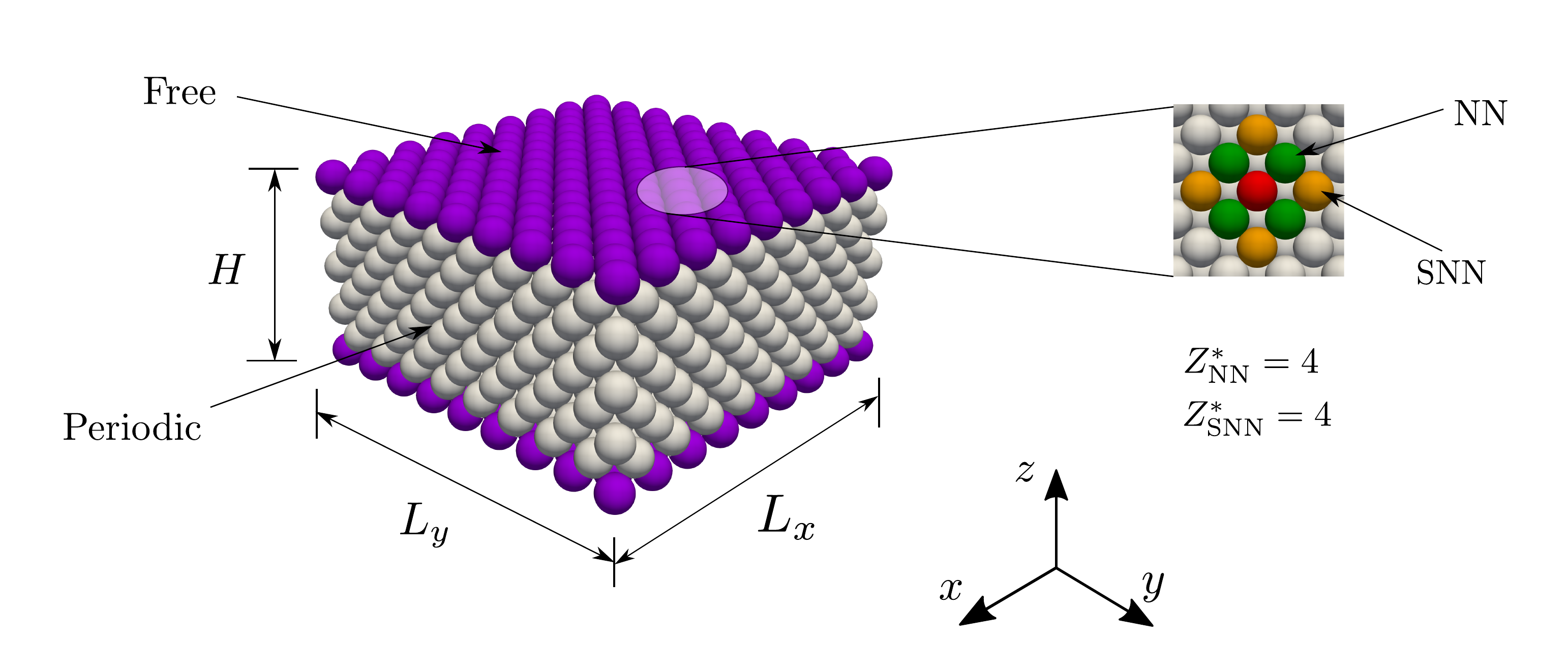,width=\textwidth}
\caption{Simulation slab used to find surface energies. Periodic boundary conditions are applied to the surface atoms on those faces perpendicular to the $x$- and $y$-directions, while atoms on the top and bottom faces (perpendicular to the $z$-direction) are free.}\label{fig:slabBox}
\end{figure}

The surface free energy density can be defined as the excess free energy $\mathcal{F}$ per unit surface area $A$, acquired by an infinite crystal when split along a plane. To simulate this scenario, we use a slab of dimensions $L_x \times L_y \times H$ with two free surfaces of area $A=L_xL_y$ separated by a height $H$ in the $z$-direction (see Fig.~\ref{fig:slabBox}). While the top and bottom surfaces are free, periodic boundary conditions are applied in the transverse $x$- and $y$-directions to simulate an infinite slab.
The reduced number of interatomic neighbors at the surfaces results in surface stresses, also referred as \textit{surface tension}. The equilibrium state of such a slab is governed by the balance of those surface stresses by the stresses in the sub-surface atomic layers \cite{streitz1994surface}. The excess free energy in the slab due to the existence of the free surfaces is retrieved by subtracting from the free energy of the slab the bulk free energy (i.e., the total energy of $N$ atoms, all having the same energy found inside the bulk of an infinite solid, where $N$ equals the number of atoms in the slab). 

This excess energy stems from two contributions: \textit{(i)}~the excess energy of atoms near the surface due to lower coordination number, and \textit{(ii)}~the strain energy induced in the sub-surface layers of atoms due to the compressive stresses imposed by the surface. One can define two different measures of the surface energy in order to capture one or both of these contributions as follows. If we subtract the bulk free energy $\tilde{\mathcal{F}}_{bulk}$ of an infinite periodic crystal in its relaxed equilibrium state from the slab's free energy (both having the same number of atoms), the obtained excess energy density, which we call $\tilde{\xi}$, is a result of both contributions \textit{(i)} and \textit{(ii)}. By contrast, to compute the surface free energy resulting solely from contribution \textit{(i)} (as for example in \cite{shenoy2005atomistic}), one must subtract the bulk energy $\hat{\mathcal{F}}_{bulk}$ of an infinite periodic crystal compressed along the $x$- and $y$-directions to impose transverse lattice spacings equal to those of the slab in equilibrium. We refer to this surface energy density as $\hat{\xi}$. This leads to two distinct definitions of free surface energy density:
\begin{equation}
    \tilde{\xi} = \frac{\mathcal{F}_{slab} - n\ \tilde{\mathcal{F}}_{bulk}}{2 A} \qquad \mathrm{and} \qquad  \hat{\xi} = \frac{\mathcal{F}_{slab} - n\ \hat{\mathcal{F}}_{bulk}}{2 A},
    \label{compute FSED}
\end{equation}
where $n$ is a factor that accounts for differences in numbers of atoms (if any) in the slab and the bulk.
Physically, we expect $\tilde{\xi}\geq\hat{\xi}$, as only $\tilde{\xi}$ accounts for the excess energy due to the compression of the inner layers. As mentioned before, the slab shown in Fig. \ref{fig:slabBox} usually tends to compress in directions transverse to the surface due to tensile surface stresses found in the metals studied here. Intuitively, one may argue that the amount of compressive strains present in equilibrium decreases as the slab thickness $H$ increases, because the inner bulk energy tends to dominate with increasing $H$. In the asymptotic limit $H\to\infty$, the transverse compression becomes negligible, so that the excess slab energy 
is only due to the reduced coordination number of the near surface atoms and
\begin{align}
    \lim_{H \rightarrow \infty} (\tilde{\xi} - \hat{\xi}) = 0 \RA \lim_{H \rightarrow \infty} \tilde{\xi} = \lim_{H \rightarrow \infty} \hat{\xi} = \bar{\xi}. \label{asymp conv}
\end{align}
Quantity $\bar{\xi}$ is the series-converged value, which can be independently obtained by using the same slab geometry of Fig.~\ref{fig:slabBox} initialized with the lattice parameter of the relaxed bulk crystal at the temperature of interest and preventing any relaxation of $L_x$ and $L_y$. If the slab is sufficiently thick to avoid interactions between the two free surfaces, \eqref{compute FSED} can be used to efficiently compute the asymptotic value $\bar{\xi}$ from a finite-size slab. Results illustrating the convergence of $\tilde{\xi}$ and $\hat{\xi}$ to $\bar{\xi}$ with increasing $H$ will be presented in Section~\ref{Thickness convergence}. 

To compute surface stresses and elastic constants, we strain the slab in the $x$- and $y$-directions, considering the aforementioned asymptotic geometry as the unstrained reference state. We impose a homogeneous deformation gradients of the form $\boldsymbol{F} = \boldsymbol{I} + \nabla\boldsymbol{u}$ 
onto the reference state of the slab ($\bfu$ denoting the 2D in-plane displacement field). From the deformation gradient, we compute the infinitesimal strain tensor as $\bfepsilon=\frac{1}{2}(\nabla\bfu+[\nabla\bfu]\T)$. To extract the surface free energy density landscape as a function of the infinitesimal strain tensor components, we simultaneously consider a periodic bulk crystal initialized with the relaxed lattice spacing at the temperature of interest, strained in exactly the same manner as the slab, and allowed to relax only in the $z$-direction, while keeping $L_x$ and $L_y$ fixed after each straining step. Subtracting the thus-obtained bulk free energy from the slab free energy at the same deformed state (possibly adjusting for different numbers of atoms in both) yields the surface free energy density landscape as
\begin{equation}
    \bar{\xi}(\bfepsilon) = \frac{\mathcal{F}_{slab}(\bfepsilon) - n\ \mathcal{F}_{bulk}(\bfepsilon)}{2A(\bfepsilon)}.
    \label{compute FSED def}
\end{equation}

Setting $\bfepsilon=\bfnull$ in \eqref{compute FSED def} recovers the infinite-thickness limit of \eqref{asymp conv}. 
Therefore, applying \eqref{compute FSED def} to a slab of sufficient thickness $H$ to avoid interactions of the two slab surfaces allows us to compute the landscape $\xi(\boldsymbol{\varepsilon})$ by repeated simulations with different strains $\boldsymbol{\varepsilon}$. Assuming that this landscape is continuous and sufficiently smooth, it can be used to derive surface stresses and elastic constants. Following \citet{shenoy2005atomistic}, the surface stress $\tau_{ij}$ and elasticity tensor $C_{ijkl}$ in the reference state can be expressed as
\begin{align}
    \tau_{ij} = \left[\bar{\xi} \delta_{ij} + \frac{\partial \bar{\xi}}{\partial \varepsilon_{ij}}  \right]_{\boldsymbol{\varepsilon}=\textbf{0}}\qquad \mathrm{and} \qquad C_{ijkl} = \left[ 2\bar{\xi}\delta_{ik}\delta_{jl} + \delta_{ij} \frac{\partial \bar{\xi}}{\partial \varepsilon_{kl}} + \frac{\partial^2\bar{\xi}}{\partial \varepsilon_{ij}\partial \varepsilon_{kl}} \right]_{\boldsymbol{\varepsilon}=\textbf{0}},
    \label{surface stress and elastic constants}
\end{align}
with indices $i,j,k,l=1,2$, where $1$ and $2$ correspond to the $x$- and $y$-directions, respectively.

The specific FCC and BCC materials chosen in this study, along with the analyzed surface orientations, are summarized in Table~\ref{Introductory Table Materials}. That table also contains the interatomic potentials used, which are all of the family of the Embedded Atom Method (EAM). The slab dimensions used in our simulations (in terms of the lattice parameter $a$ of each material) are provided in Table~\ref{Slab dimensions table}.

\begin{table}[h!]
\centering
\begin{tabular}{| m{1.0cm} |  m{1.0cm} | m{1.0cm}| m{1.0cm} |  m{1.0cm} | m{1.0cm}| m{1.0cm} |  m{1.2cm} |  m{3.0cm} | } 
\hline
 & \multicolumn{3}{c|}{Surface Symmetry Group} & \multicolumn{3}{c|}{ Number of Neighbors} & \multirow{3}{1.2cm}{Material} & \multirow{3}{1.2cm}{Potential}   \\
\cline{2-7} & (001) & (011) & (111) & (001) & (011) & (111)  & &   \\
  \hline
  \multirow{3}{1cm}{FCC} & \multirow{3}{1.0cm}{4mm} & \multirow{3}{1.0cm}{2mm} & \multirow{3}{1.0cm}{6mm} & \multirow{3}{1.0cm}{\{4,4\}} & \multirow{3}{1.0cm}{\{2,2\}} & \multirow{3}{1.0cm}{\{6,0\}} &  $\text{Cu}^*$ & \citet{mishin2001structural}  \\
 & & & & & & &  Ni  & \citet{sheng2011highly}  \\
 & & &  & & &  &  Al & \citet{liu2004aluminium}  \\
\hline
\multirow{3}{1cm}{BCC} & \multirow{3}{1.0cm}{4mm} & \multirow{3}{1.0cm}{2mm} & \multirow{3}{1.0cm}{3m} & \multirow{3}{1.0cm}{\{0,4\}} & \multirow{3}{1.0cm}{\{4,2\}} & \multirow{3}{1.0cm}{\{0,0\}}   & $\text{Fe}^*$ &  \citet{chamati2006embedded}  \\
  & & & & & & & W & \citet{marinica2013interatomic} \\
  & & &  & & & & V & \citet{olsson2009semi} \\
\hline
\end{tabular}
\captionof{table}{Different material surfaces being investigated in this study with their respective interatomic potentials and crystallographic surface orientations. The given numbers of neighbors $\{Z^*_{NN},Z^*_{SNN}\}$ for each surface type represent the number of nearest $(Z^*_{NN})$ and second nearest $(Z^*_{SNN})$ neighbors present for any atom on the surface, along with the symmetry group of each surface. (Asterisks indicate those materials for which results will be validated through LAMMPS simulations.)}
\label{Introductory Table Materials}
\end{table}

\begin{table}[h!]
\centering
\begin{tabular}{| m{1.0cm} | m{1.0cm} |  m{1.0cm} | m{1.0cm} | m{1.3cm} |  m{1.0cm} |  m{1.0cm} | m{1.0cm} | m{1.3cm} |} \hline & \multicolumn{4}{c|}{FCC} & \multicolumn{4}{c|}{BCC} \\
\hline & \multirow{2}{1.0cm}{$L_x$} & \multirow{2}{1.0cm}{$L_y$} & \multicolumn{2}{c|}{$H$}  & \multirow{2}{1.0cm}{$L_x$} & \multirow{2}{1.0cm}{$L_y$} & \multicolumn{2}{c|}{\textbf{$H$}}  \\
\cline{4-5} \cline{8-9} &  &  & $H_{\text{min}}$ & $H_{\text{max}}$  &  &  & $H_{\text{min}}$ & $H_{\text{max}}$ \\ 
\hline (001) & $8a$  & $8a$  & $4a$  & $180a$   & $8a$  & $8a$  & $4a$  & $180a$  \\ \hline (011)
 & $8a$  & $6\sqrt{2} a$  & $3\sqrt{2}a$  & $125\sqrt{2}a$   & $8a$  & $6\sqrt{2}a$  & $2\sqrt{2}a$  & $130\sqrt{2}a$  \\ \hline
(111) & $6\sqrt{2}a$  & $3\sqrt{6}a$  & $2\sqrt{3}a$   & $104\sqrt{3}a$    & $6\sqrt{2}a$  & $4\sqrt{6}a$   & $3\sqrt{3}a$   & $104\sqrt{3}a$ \\
\hline 
\end{tabular}
\captionof{table}{Slab dimensions used in simulations for different crystallographic orientations in terms of the lattice parameter $a$. }
\label{Slab dimensions table}
\end{table}

\subsection{GPP simulation details} 
\label{GPP simulation details}

We consider three FCC metals (Cu, Ni, Al) and three BCC metals (Fe,W,V), each with three different crystallographic surface orientations, as summarized in Table~\ref{Introductory Table Materials}. The slab dimensions used in simulations (see Fig \ref{fig:slabBox}) are provided in Table~\ref{Slab dimensions table} for all three surface orientations. For each case, we simulate thicknesses ranging from $H_{\text{min}}$ to $H_{\text{max}}$ and seek convergence of $\tilde{\xi}$ and $\hat{\xi}$ with increasing $H$ (see Section~\ref{Thickness convergence}). For each configuration, we solve Eqs.~\eqref{EOM quasistatic} at a fixed uniform temperature $T$ in a quasistatic fashion to find the mean atomic positions $\bar\bfq$ and position variances $\Sigma$ of all atoms in the slab.
To obtain the surface free energy density landscape $\bar{\xi}(\boldsymbol{\varepsilon})$, we strain the slab geometry (and the corresponding bulk) in the nine independent straining modes summarized in Table~\ref{Loading cases table}. Specifically, we apply deformation gradients $\bfF=\id+\epsilon \nabla \bfu$, where $\nabla \bfu$ takes one of the displacement gradient forms of Table~\ref{Loading cases table} (cf.~\citep{sievers2020computational}) , and $\epsilon$ is a scalar incremented in simulations from $-0.0015$ to $0.0015$ in steps of $0.0001$. Each deformation case is applied to the slab, so its periodic faces in the $x$- and $y$-directions are also deformed according to the deformation gradient. For each material and surface orientation, we numerically map the energy landscape $\bar{\xi}(\boldsymbol{\varepsilon})$ through a large set of simulations. The derivatives in \eqref{surface stress and elastic constants} are computed by first fitting a fourth-order multivariate polynomial function to the surface free energy landscape $\bar{\xi}(\boldsymbol{\epsilon})$, followed by differentiation of that fitting function. 

\begin{center}
\begin{table}[h!]
\centering
\begin{tabular}{| m{2.5cm} | m{3.0cm} ||  m{2.5cm} | m{3.0cm} |} \hline \multicolumn{2}{|c||}{Simple} & \multicolumn{2}{c|}{Combinations} \\
\hline \multirow{2}{3.0cm}{Uniaxial} & $\bfe_1 \otimes \bfe_1$ & Biaxial Opposite  & $\bfe_1 \otimes \bfe_1 - \bfe_2 \otimes \bfe_2$ \\ \cline{3-4}  & $\bfe_2 \otimes \bfe_2$ & \multirow{4}{3.0cm}{Shear + Uniaxial}  & $\bfe_1 \otimes \bfe_2 + \bfe_1 \otimes \bfe_1$ \\ \cline{1-2} \multirow{2}{3.0cm}{Biaxial} & \multirow{2}{3.0cm}{$\bfe_1 \otimes \bfe_1 + \bfe_2 \otimes \bfe_2$} & & $\bfe_1 \otimes \bfe_2 - \bfe_1 \otimes \bfe_1$ \\  &  & & $\bfe_1 \otimes \bfe_2 + \bfe_2 \otimes \bfe_2$ \\ \cline{1-2} Shear & $\bfe_1 \otimes \bfe_2$ & & $\bfe_1 \otimes \bfe_2 + \bfe_2 \otimes \bfe_2$ \\ \hline
\end{tabular}
\caption{Displacement gradients $\nabla \bfu$ used in the deformation gradient $\bfF=\id+\epsilon \nabla \bfu$ for the different load cases to approximate the free surface energy density landscape. $\bfe_1$ and $\bfe_2$ are orthogonal unit vectors in the $x$- and $y$-directions.}
\label{Loading cases table}
\end{table}
\end{center}


\subsection{MD simulation details} 
\label{TI}
To validate the surface free energies obtained from the GPP formulation, analogous MD simulations have been performed with the same interatomic potentials for FCC copper and BCC iron. We here provide a brief summary of the thermodynamic integration in LAMMPS, which we used to compute free energies from MD simulations. Thermodynamic integration is a general name given to a class of methods to find free energies in atomistic simulations. This technique can be used in both \textit{ab initio} \cite{malica2020quasi} or potential based MD simulations \cite{grochola2002simulation,frolov2009temperature,frenkel1984new}. However, we elaborate here on its implementation within the framework of MD simulations. Starting from a known equilibrium state whose free energy is known \textit{a priori}, a quasistatic reversible path is constructed, which takes the system from the known free energy state to an equilibrium state of interest (e.g., the relaxed equilibrium of a slab at finite temperature). The work done along this path is the difference between the free energies of the equilibrium state of interest and the known initial state. The choice of the initial state depends on the application of interest. For example, \citet{grochola2002simulation} found the excess free energy of a surface by switching from a bulk Hamiltonian to a slab Hamiltonian. Note that one does not need to initially know the absolute free bulk energy here, because the surface free energy is the difference between the slab and bulk free energies. This process, however, must be repeated for every temperature of interest, thus implying significant computational expenses and efforts for the present study. As an alternative, \citet{frolov2009temperature} used the quasi-harmonic crystal approximation at some fixed temperature as the starting point and performed a path integration parameterized by temperature to obtain the surface free energy at other temperatures. This class of path integration (or thermodynamic integration) methods was first introduced by \citet{frenkel1984new}. The basic idea was to consider an Einstein approximation as the initial equilibrium state. Considering the system to be a canonical ($NVT$) ensemble and the thermodynamic path parameterized by a variable $\lambda$, the free energy is given as
\begin{align}
    \mathcal{F}(N,V,T;\lambda)  = -k_B T\, \mathrm{ln} Z(N,V,T;\lambda), 
    \label{free energy canonical}
\end{align}
where $Z(N,V,T;\lambda)$ is the partition function, given by the phase average
\begin{align}
    Z(N,V,T;\lambda) = \int \frac{\exp[-\mathcal{H}(\boldsymbol{z};\lambda)/k_B T]}{N! h^{3N} } \dd\bfz,
\end{align}
where integration over all of phase space is implied, with $\calH$ being the total Hamiltonian. Suppose that $\mathcal{H}_I$ and $\mathcal{H}_F$ denote, respectively, the initial (Einstein approximation) and final Hamiltonians. Our switching path is parametrized by $\mathcal{H} = (1-\lambda) \mathcal{H}_I + \lambda \mathcal{H}_F $, so that changing $\lambda$ from $0$ to $1$ marks the transition from the initial to the final state. For fixed values of $N,\,V$ and $T$, \eqref{free energy canonical} thus yields
\begin{align}
    \mathrm{d} \mathcal{F} = \Big\langle \frac{ \mathrm{d} \mathcal{H} }{ \mathrm{d} \lambda } \Big\rangle \, \mathrm{d} \lambda = \langle \mathcal{H}_F - \mathcal{H}_I \rangle \, \mathrm{d} \lambda.
    \label{TI dF}
\end{align}

In equilibrium thermodynamic integration methods, the phase average on the right-hand side of Eq.~\eqref{TI dF} is computed at discrete values of $\lambda$, and the equation is integrated numerically. This requires an individual equilibrium simulation at each discrete value of $\lambda$ during a single temperature integration. \citet{freitas2016nonequilibrium} have proposed a quicker procedure by smoothly varying $\lambda$ from $0$ to $1$ in a single non-equilibrium simulation. Since this process is not quasistatic, the work done, $W_{I\rightarrow F}$, includes an extra dissipative term. Assuming this dissipative contribution to be the same for the forward (i.e., initial to final, or $I \rightarrow F$) and backward ($F \rightarrow I$) processes, it can be eliminated by writing the actual work done as $(W_{I\rightarrow F}-W_{F\rightarrow I})/2$. LAMMPS implements this procedure using the \textit{fix ti/spring} command. It equilibrates the system for a specified time $t^{eq}$ in the final state, switches to the initial state in a specified time $t^{sw}$, equilibrates again for the same time and switches back to the final state. This two-way switching allows to compute $W_{F\rightarrow I}$ and $W_{I\rightarrow F}$ and thus the desired actual free energy. 

The initial state (Einstein approximation) is a system of independent harmonic oscillators centered about their initial lattice positions $\bfq_i^0$. The Hamiltonian of such a system is
\begin{align}
    \mathcal{H}_I = \sum_{i=1}^N \left( \frac{|\bfp_i|^2}{2m_i} + \frac{k_i}{2} \left|\bfq_i-\bfq_i^0\right|^2 \right).
\end{align}
Using this definition of the Hamiltonian in a canonical ensemble probability distribution and the relation $k_i = 3k_B T/\langle \vert\Delta \bfq\vert^2 \rangle$ (with $\Delta \bfq=\bfq_i-\bfq_i^0$) yields the independent GPP distribution function~\eqref{independent GPP}. Unlike the GPP approach, close estimates of the mean squared displacements are determined here using prior bulk simulations. These estimates define the starting state for performing the Frenkel-Ladd thermodynamic integration on our system.

In a prior simulation, we obtain the relaxed bulk lattice parameters and mean-squared displacements $\langle \vert\Delta \bfq\vert^2 \rangle$, using the isoenthalpic-isobaric ($NPH$) ensemble and a Langevin thermostat for five equispaced temperature levels from $100$~K to $500$~K. The thus-obtained lattice parameters are then used to create bulk and slab geometries. The slab is given the boundary conditions discussed in Section~\ref{sec:Configuration} (see Fig.~\ref{fig:slabBox}), and it is allowed to relax under an $NVT$ ensemble sampled with a Langevin thermostat. Large simulation cells with $20\times 20\times 20$ unit cells are used to approach the thermodynamic limit and to ensure the convergence of thermodynamic averages with increasing system size. The bulk models have 16,000 and 32,000 atoms, respectively, for the BCC and FCC lattice types. While doing the Frenkel-Ladd path integration, the previously computed $\langle\vert\Delta \bfq\vert^2\rangle$ values are used to find the spring constants for the Einstein crystal approximation. Long equilibration times of $10$~ns are used for the slabs to obtain accurate time averages, while the thermodynamic switching between the original system and the quasi-harmonic approximation is performed in $2$~ns.

\section{Results and Discussion} 
\label{results}

In this section, we summarize the results of our GPP methodology for surface free energies, surface stresses, and elastic constants as functions of temperature. In Section~\ref{Thickness convergence} we begin by demonstrating convergence of the lattice spacing and the surface free energies $\tilde{\xi}$ and $\hat{\xi}$ with increasing slab thickness. In the following we focus on the converged infinite-thickness scenario, for which we report the effect of temperature on the surface properties -- in comparison with LAMMPS data for selected materials for validation. Having validated the GPP methodology, we present in Section~\ref{surface elastic constants results} results for surface stresses and surface elastic constants vs.\ temperature.

\subsection{Thickness convergence study}
\label{Thickness convergence}

\begin{figure}[t!]
\centering
\includegraphics[width = \textwidth]{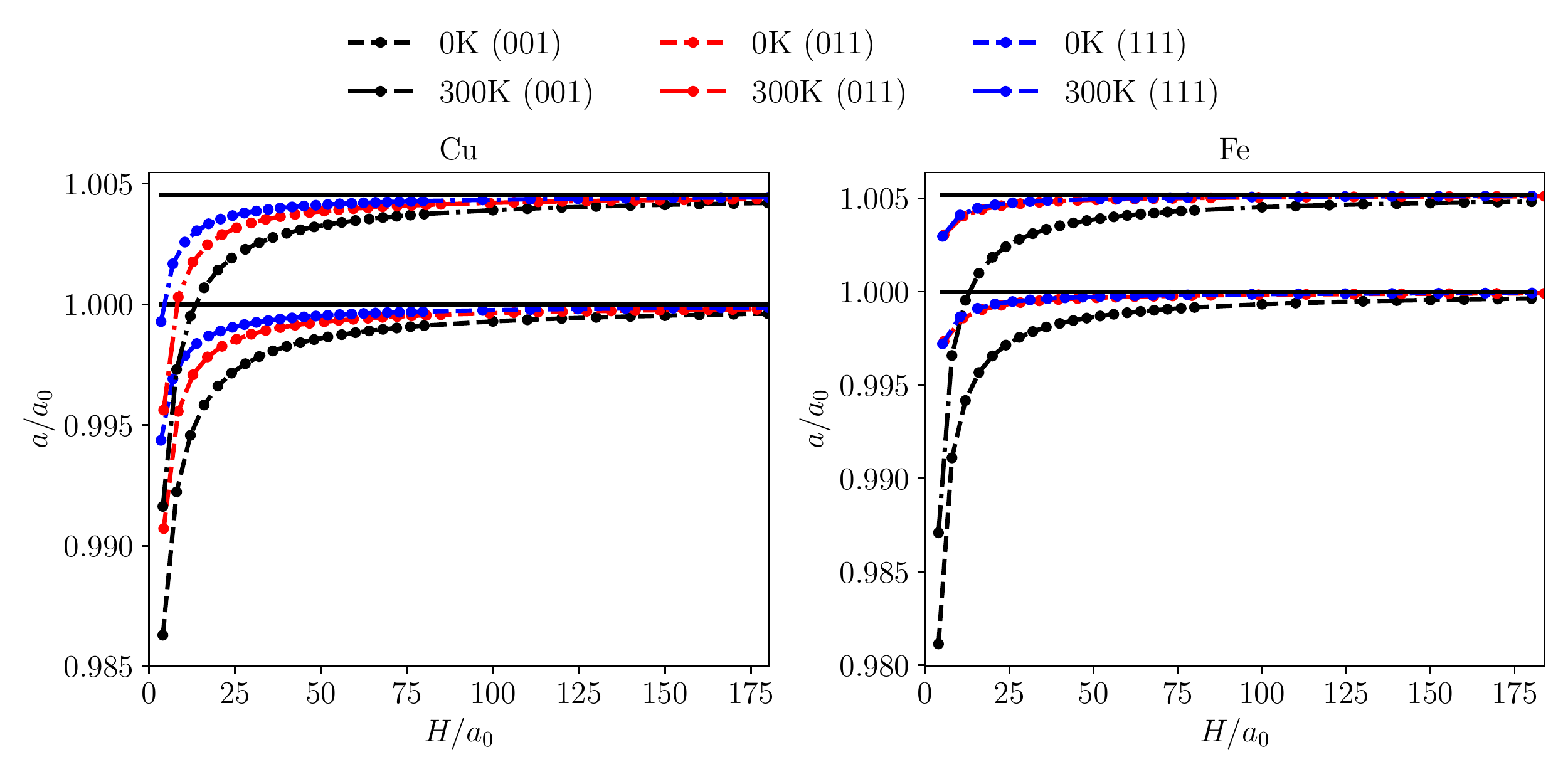}
\caption{Lattice parameter convergence with increasing thickness for copper and iron slabs at 0~K and 300~K (normalized by the respective bulk lattice parameter $a_0$ of each material). The corresponding asymptotic values are shown as black solid lines.}\label{fig:lattice parameter convergence}
\end{figure}

In this Section, we verify the convergence of $\tilde{\xi}$ and $\hat{\xi}$ with increasing slab thickness $H$. Fig.~\ref{fig:lattice parameter convergence} shows the convergence of the lattice parameter $a$ in the $x$-$y$-plane at 0~K and 300~K for copper and iron slabs (each normalized by the respective lattice parameter $a_0$ in the bulk). In all cases, the lattice parameter converges to the correct bulk value, as expected. The convergence for copper is fastest for the (111) surface, while the values for iron converge fastest for the (011) surface. The (001) surface shows slowest convergence for both materials. As expected, the compression of the sub-surface atomic layers vanishes as the slab thickness approaches infinity, leading to a uniform bulk spacing in the $x$-$y$-plane. As expected, the surface lattice parameter increases with temperature (as does the bulk lattice parameter).

\begin{figure}[!ht]
\centering
\includegraphics[width = \textwidth]{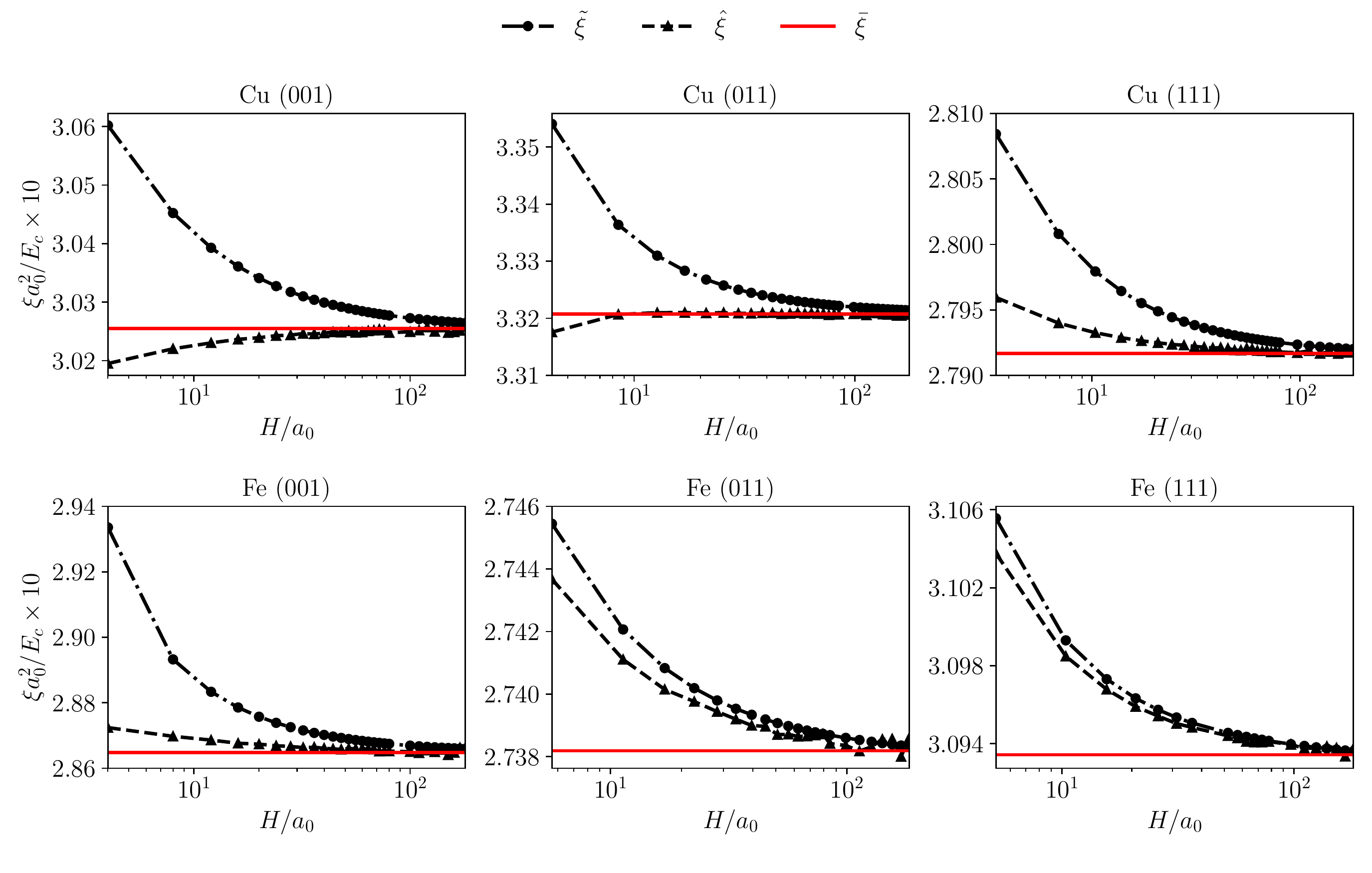}
\caption{Convergence of the two surface free energy density measures $\tilde{\xi}$ and $\hat{\xi}$ to the asymptotic value $\bar{\xi}$ with increasing slab thickness at 300~K. Slab thickness $H$ is normalized by the lattice spacing $a_0$, while the surface energy density $\xi$ is normalized by the zero-temperature cohesive energy $E_c$.
}\label{fig:FSED convergence}
\end{figure}

The asymptotic convergence in Eq.~\eqref{asymp conv} is visualized in Fig.~\ref{fig:FSED convergence}, which plots the surface free energies for different surface orientations of copper and iron at 300~K. For all surfaces, we consistently observe $\tilde{\xi}>\hat{\xi}$ due to the additional strain energy of the sub-surface atomic layers. The convergence characteristics observed in Fig.~\ref{fig:FSED convergence} can be related to the lattice parameters of Fig.~\ref{fig:lattice parameter convergence}. For example, slab surfaces such as (011) and (111) in iron, which exhibit little compression of the sub-surface layers, show little difference  between $\tilde{\xi}$ and $\hat{\xi}$. The other surfaces show a significant compression in the $x$-$y$-plane and therefore considerable differences between the values of $\tilde{\xi}$ and $\hat{\xi}$ for small slab thicknesses.

\begin{figure}
\centering
\includegraphics[width = \textwidth]{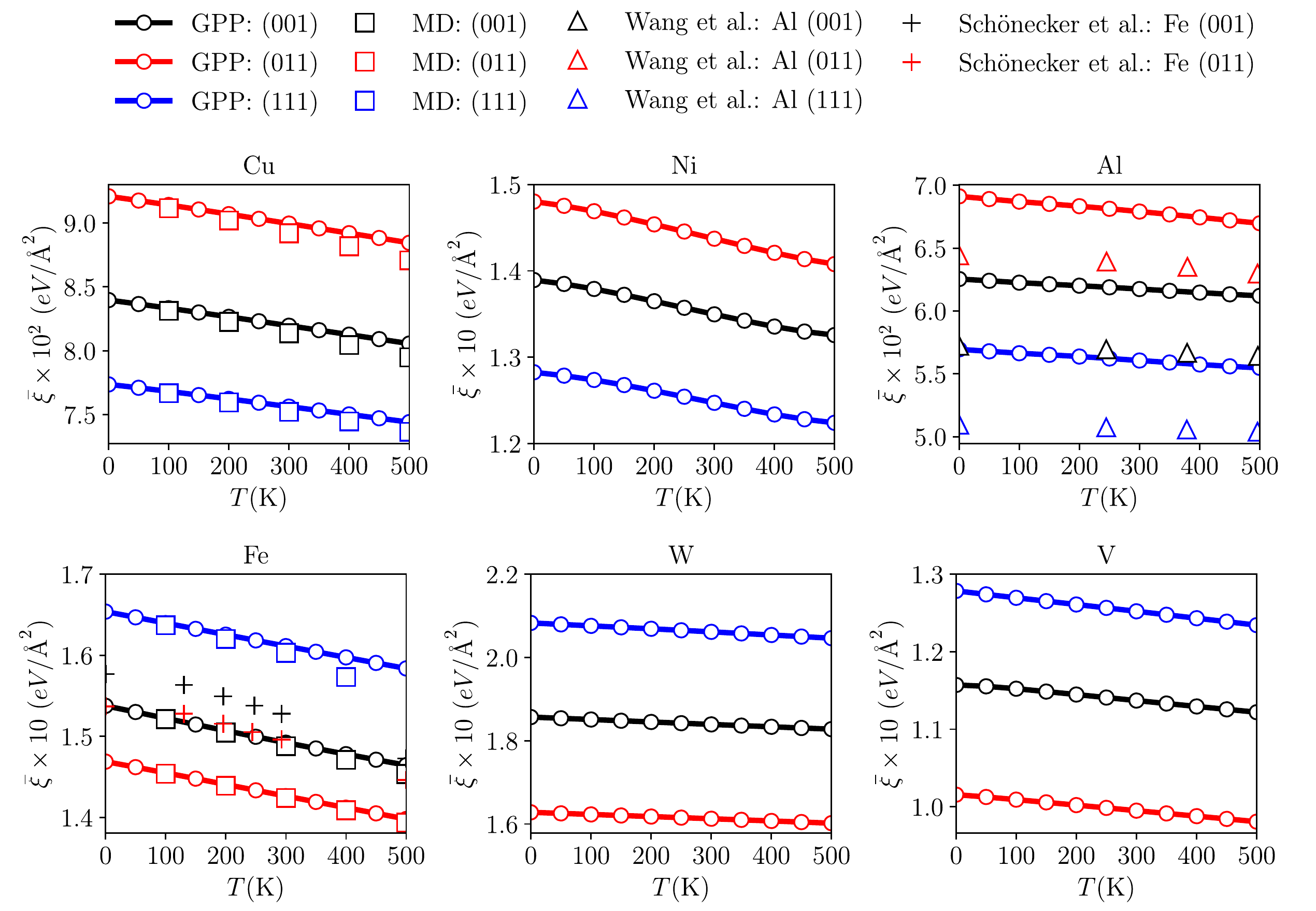}
\put(-470,145){$(a)$}
\put(-470,5){$(b)$}
\caption{ Surface free energy density $\bar{\xi}(T)$ vs,\ temperature for (a) FCC and (b) BCC metals, in comparison to MD data and literature values (\citet{wang2019effects,schonecker2015thermal})
obtained from Density Functional Theory (DFT).}\label{fig:FSED}
\end{figure}

Fig.~\ref{fig:FSED} illustrates the dependence of the surface free energy density on temperature for all studied FCC and BCC metal surfaces. (Table~\ref{Surface free energies table} lists the values of $\bar{\xi}$, as obtained from the GPP framework for every metal and surface orientation at four temperature levels.) A unanimous trend for all materials, $\bar{\xi}$ decreases with temperature. For FCC metals we observe $\bar{\xi}_{(011)}>\bar{\xi}_{(001)}>\bar{\xi}_{(111)}$, which can be attributed to the number of nearest and second-nearest neighbor bonds that are broken by creating the respective surface (cf.~Table~\ref{Introductory Table Materials} for the number of nearest and second-nearest neighbors, $Z^*_{NN}$ and $Z^*_{SNN}$, of a surface atom). Although an FCC (001) surface has most neighbors available, the FCC (111) face has most nearest neighbors. This is why it has the least surface energy, followed by the (001) and finally (011) surface, the latter having the least number of neighbors. By contrast, the BCC (111) surface has neither a nearest nor a second-nearest neighbor on the surface. The (001) surface has only second-nearest neighbors and the (011) surface has both nearest and second-nearest neighbors on the surface. Therefore, the trend for BCC metals is $\bar{\xi}_{(111)}>\bar{\xi}_{(001)}>\bar{\xi}_{(011)}$.

The values of $\bar{\xi}$ obtained using thermodynamic integration within LAMMPS match closely with the reported data obtained from our GPP framework. Discrepancies between GPP and MD data increase with increasing temperature. This is expected since the quasi-harmonic assumption of independent Gaussian phase packets for every atom becomes a poorer approximation of the actual crystal due to stronger anharmonic interactions in lieu of more atomic vibrations at higher temperatures. However, the GPP results always stay within 2\% of the LAMMPS values. This comparison serves as a validation for using the GPP-based approach to study surface free energies for the aforementioned temperature range -- which is why we will use the GPP-based approach in the following for our results. We note that the Fe (111) surface, which is the least accurate with increasing temperature, is a special case, as it tends to show faceting at elevated temperature: surface atoms may hop out of the surface, creating vacancies, as shown by \citet{grochola2002new}, who used a specifically adapted thermodynamic integration technique. The usual Frenkel-Ladd path fails here because the atoms move too far from their initial positions, about which the Einstein solid approximation is made. Therefore, we report MD results for this surface only up to 400K. The shown results are also in good agreement with the MD finite temperature calculations for Fe by \citet{xing2020temperature} and for Cu by \citet{frolov2009temperature}.
For reference, Fig.~\ref{fig:FSED} also contains data from Density Functional Theory (DFT) for Fe by \citet{schonecker2015thermal} and for Al by \citet{wang2019effects}. Although the DFT values come with an offset from those reported here from MD and the GPP approach, their trends with temperature match very well. Note that we would not expect perfect agreement between ab-initio data and those from fitted MD potentials. Most importantly, we see excellent agreement between MD and our GPP approach (using the same interatomic potentials) across the investigated temperature range, while the GPP framework is computationally significantly less expensive (replacing MD time-stepping by a quasistatic relaxation), less cumbersome (replacing thermodynamic integration by a simple optimization problem), more robust (not depending on time step size, averaging times, etc.), and easily reproducible.

\begin{center}
\begin{table}[h!]
\begin{tabu}{| m{1.0cm} | m{1.0cm} |  m{1.0cm} | m{1.0cm} | m{1.0cm} |  m{1.0cm} |  m{1.0cm} | m{1.0cm} | m{1.0cm} | m{1.0cm} |  } \tabucline{1-10}
& \multicolumn{3}{c|}{Cu} & \multicolumn{3}{c|}{Ni} & \multicolumn{3}{c|}{Al} \\ \cline{2-10}
& (001) & (011) & (111) & (001) & (011) & (111) & (001) & (011) & (111) \\ \tabucline{1-10}
0~K & 0.0840 & 0.0921 & 0.0774 & 0.1389 & 0.1480 & 0.1283 & 0.0625 & 0.0691 & 0.0569 \\ \hline
100~K & 0.0833 & 0.0914 & 0.0768 & 0.1379 & 0.1469 & 0.1274 & 0.0623 & 0.0687 & 0.0566 \\ \hline
300~K & 0.0820 & 0.0900 & 0.0756 & 0.1350 & 0.1437 & 0.1247 & 0.0617 & 0.0679 & 0.0561 \\ \hline
500~K & 0.0806 & 0.0884 & 0.0744 & 0.1326 & 0.1408 & 0.1224 & 0.0612 & 0.0670 & 0.0555 \\ \tabucline{1-10}\tabucline{1-10}
& \multicolumn{3}{c|}{Fe} & \multicolumn{3}{c|}{W} & \multicolumn{3}{c|}{V}  \\ \cline{2-10}
& (001) & (011) & (111) & (001) & (011) & (111) & (001) & (011) & (111) \\ \tabucline{1-10}
0~K & 0.1538 & 0.1469 & 0.1654 & 0.1857 & 0.1628 & 0.2083 & 0.1157 & 0.1016 & 0.1279  \\ \hline
100~K & 0.1522 & 0.1455 & 0.1640 & 0.1851 & 0.1623 & 0.2076 & 0.1152 & 0.1009 & 0.1270 \\ \hline
300~K & 0.1492 & 0.1426 & 0.1611 & 0.1839 & 0.1613 & 0.2062 & 0.1137 & 0.0995 & 0.1252 \\ \hline
500~K & 0.1464 & 0.1398 & 0.1584 & 0.1828 & 0.1602 & 0.2046 & 0.1122 & 0.0981 & 0.1234 \\ \tabucline{1-10}
\end{tabu}

\caption{Surface free energy densities $\bar{\xi}$ for all metals and surface orientations studied here. All values are in $eV/$\AA$^2$.}
\label{Surface free energies table}
\end{table}
\end{center}

\subsection{Isothermal surface stresses and surface elastic constants}
\label{surface elastic constants results}

Following the procedure outlined in Section~\ref{methodology}, we here report the variations of surface stresses and surface elastic constants, as obtained  from the GPP approach (and validated by reference MD calculations). Section~\ref{sec:FCC} summarizes all FCC metal data, followed by the BCC metals in Section~\ref{sec:BCC}. We start with the variation of the average surface stress $\tau_{\mathrm{avg}}=(\tau_{11}+\tau_{22})/2)$ vs.\ temperature, including a comparison with MD results for Cu and Fe. For visualizing the surface elastic tensor, we present the directional compliance $S$ in a polar plot for different surfaces. The directional compliance as a function of the direction $\boldsymbol{d}$ can be computed as
\begin{align}
    S(\boldsymbol{d}) = C^{-1}_{ijkl}d_i d_j d_k d_l, \qquad\text{where}\quad |\bfd|=1
\end{align}
and we use Einstein's summation convention over repeated indices.
The directional stiffness $E(\bfd)=1/S(\bfd)$ is the inverse of $S$. However, since the surface elastic tensor $\boldsymbol{C}$ is not necessarily positive definite (as also emphasized by \citet{shenoy2005atomistic}), the value of $S$ may changes sign while traversing along the polar axis. As this creates large jumps in the directional stiffness, we rather plot $S=1/E$ in the polar plots. We stress that a lack of positive-definiteness in the \textit{surface} elastic constants (unlike the \textit{bulk} elastic constants) does not necessarily constitute a material instability. In fact, all slabs investigated here were stable under the given boundary conditions. This is due to the sub-surface atoms in the slab, whose potential energy may stabilize a surface with non-positive-definite elastic constants. It is interesting to note that this also leads to the surface's directional Poisson's ratio to lie outside the usual range of $[-1,0.5]$. 

As only limited literature data is available for finite-temperature surface stresses and elastic constants in metals, we usually report comparisons of elastic constants for 0~K and, if available, for stresses at finite temperature.

\subsubsection{FCC metals}
\label{sec:FCC}

\begin{figure}
\centering
\includegraphics[width = \textwidth]{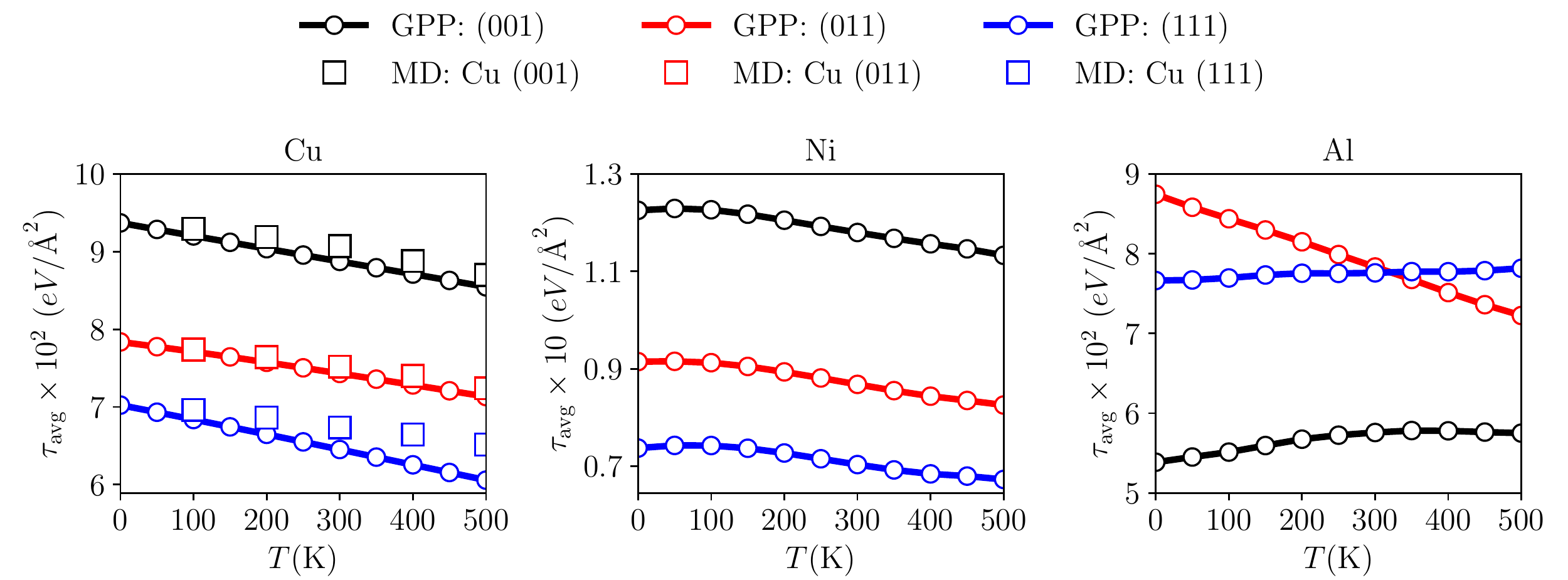}
\caption{Average surface stress vs.\ temperature for FCC metals (MD data for Cu, obtained via thermodynamic integration in LAMMPS, is included for comparison).}\label{fig: stresses FCC}
\end{figure}

Fig.~\ref{fig: stresses FCC} summarizes surface stresses of FCC metals vs.\ temperature. For Cu, we again observe a good general agreement between our GPP data and those obtained from thermodynamic integration via MD. A common decreasing trend is observed with increasing temperature for all surfaces, except for the (001) and (111) surfaces of Al. Ni displays a change in slope, initially stresses increase by roughly $0.7\%$ for the (111) surface and by less than $0.2\%$ for the other surfaces, before they start to decrease. The effect of temperature on the surface stress is minimal for Ni and maximal for Al, which is expected since it is inversely proportional to the melting points of the metals. Although not shown here, the Monte-Carlo values obtained by \citet{frolov2009temperature} for (011) Cu, using the same interatomic potential, match the GPP results to within $1\%$ relative error. The overall trend for FCC metal surfaces is $\tau^{\text{avg}}_{(001)}>\tau^{\text{avg}}_{(011)}>\tau^{\text{avg}}_{(111)}$ for Cu and Ni at all temperatures. For Al, the trend is $\tau^{\text{avg}}_{(011)}>\tau^{\text{avg}}_{(111)}>\tau^{\text{avg}}_{(001)}$ below 300~K, above which $\tau^{\text{avg}}_{(111)}>\tau^{\text{avg}}_{(011)}$. The 0~K trend for Al, as shown in \cite{shenoy2005atomistic} using the potential developed by \citet{liu2004aluminium}, matches our trend, while using the potential by \citet{voter1994intermetallic} shows a trend similar to Cu and Ni in our work.

\begin{figure}
\centering
\includegraphics[width = \textwidth]{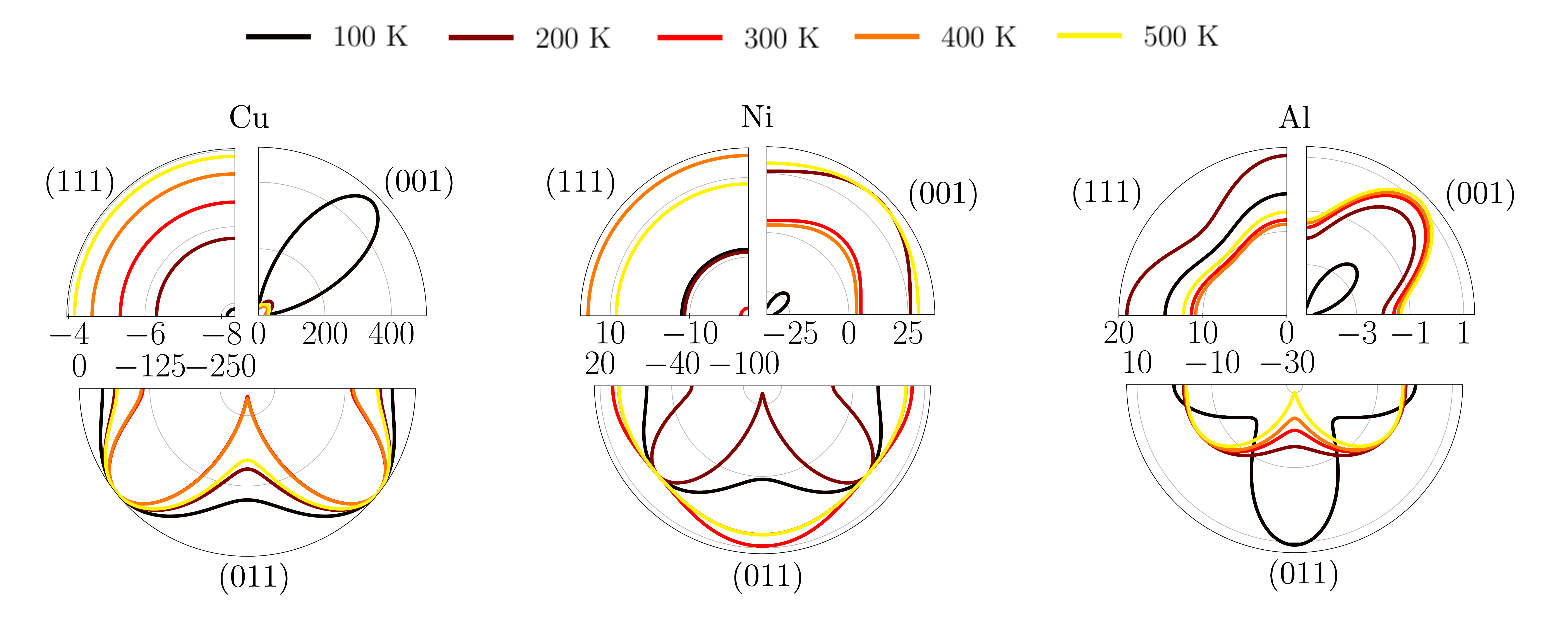}
\caption{Polar compliance plots for FCC metals. Per the FCC surface symmetries, only one quarter/semi-circle of the full polar plot is shown for different surface orientations, which can be rotated about the origin to visualize the full polar plots (see Table~\ref{Introductory Table Materials}). All compliance values are in \AA$^2/eV$.\label{fig: Polar plot FCC}}
\end{figure}

Fig.~\ref{fig: Polar plot FCC} shows the directional compliance for all surface orientations as effective quarter or semi polar plots (detailed values of surface stresses and elastic constants are consolidated in Appendix~\ref{sec:FCCData}; note that minor symmetries $C_{ijkl}=C_{jikl}=C_{ijlk}$ hold, while major symmetries do not necessarily hold for surface elastic tensors). It is difficult to deduce any general trends with temperature -- even for the isotropic (111) surfaces. Yet, the following explanations can be given for surface stresses and associated elastic constants for each of the three surface orientations.

The (001) surface in FCC metals is characterized by isotropic surface stresses but anisotropic elastic constants, as expected from the 4mm symmetry of this surface. This anisotropy is clearly visible by the non-circular contours in Fig.~\ref{fig: Polar plot FCC}. Surface stresses for Cu at 0~K are higher by $8.8 \%$ than those obtained from first-principle calculations by \citet{gumbsch1991interface}, and significantly higher (by $44\%$) than the molecular statics results of \citet{shenoy2005atomistic}, who used the (older) EAM potential of \citet{oh1988simple}.  The reported $C_{1111}=C_{2222}$ values are negative, while $C_{1122}=C_{2211}$ is positive for all metals, as also seen for 0~K results obtained by \citet{shenoy2005atomistic}. At higher temperature, the $C_{1122}$ values become negative for Al.

The (011) surface has a two-fold 2mm rotational symmetry and therefore different values for $\tau_{11}$ and $\tau_{22}$. All studied FCC metals show a decreasing trend of the (011) average surface stress with temperature. Unlike for the potential used in \citet{shenoy2005atomistic}, we find a positive surface stress for Ni at 0~K. Also, the relation $\tau_{11}>\tau_{22}$ holds for all examined metals and temperatures. This is expected, as every surface atom has two nearest neighbors along the $x$-direction and second-nearest neighbors along the $y$-direction, resulting in a stronger inward surface contraction along the $x$-direction. All elastic constants are negative for this surface for all three metals. The $C_{1111}$ values are significantly larger in magnitude than $C_{2222}$, while $C_{1122}$ and $C_{2211}$ are close to each other.

The close-packed (111) plane of FCC metals is isotropic with a 6mm symmetry, so that the surface stresses and elastic constants are isotropic. The latter is verified by the identity $2C_{1212}=C_{1111}-C_{1122}$. The only material seen to deviate from this trend is Al. Note that this finding differs from  \citeauthor{shenoy2005atomistic}'s results for Al at 0~K, who used different interatomic potentials. The surface stress generally decreases with temperature, again except for Al. $C_{1111}=C_{2222}$ values are negative at all temperatures for Cu and Ni but positive for Al, as also seen in \cite{shenoy2005atomistic}. An interesting observation for Cu and Ni is that values for $C_{1111}$ and $C_{1122}$ are similar in magnitude but opposite in sign, which makes $C_{1212}$ also of similar magnitude by the above identity. Combined with the fact that these are isotropic surfaces, the above results in the surface's Poisson's ratio to be close to $-1$.

\subsubsection{BCC Metals}
\label{sec:BCC}

\begin{figure}[b!]
\centering
\epsfig{file = 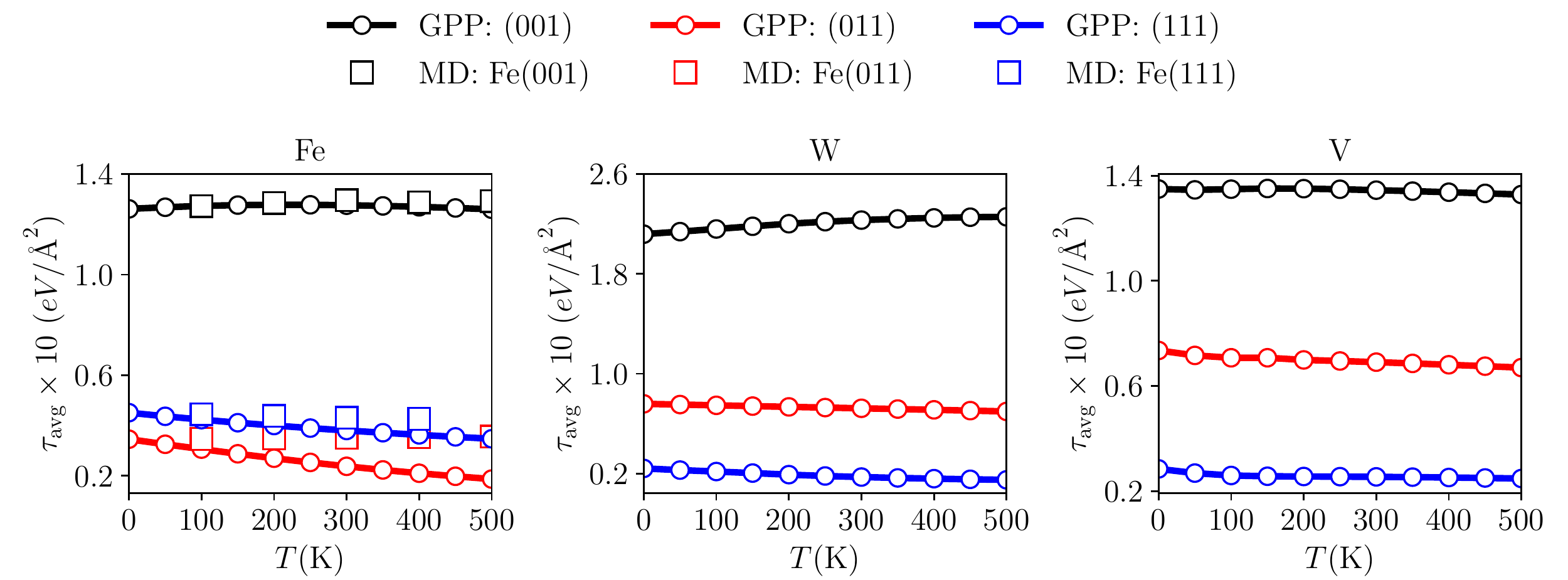,width=\textwidth}
\caption{Average surface stress vs.\ temperature for BCC metals}\label{fig: stresses BCC}
\end{figure}

\begin{figure}[ht]
\centering
\includegraphics[width = \textwidth]{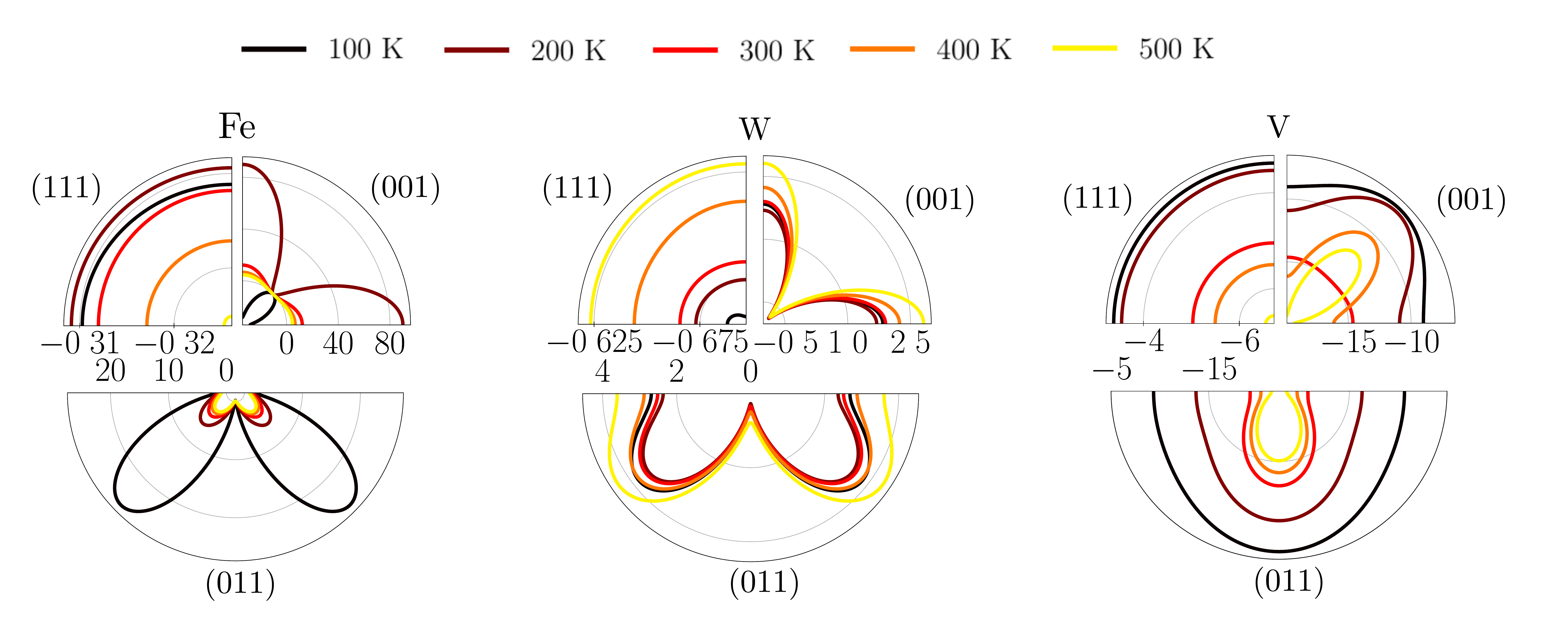}
\caption{ Polar compliance plots for BCC metals. Per the BCC surface symmetries, only one quarter/semi-circle of the full polar plot is shown for different surface orientations, which can be rotated about the origin to visualize the full polar plots (see Table~\ref{Introductory Table Materials}). All compliance values are in \AA$^2/eV$.}\label{fig: Polar plot BCC}
\end{figure}

Fig.~\ref{fig: stresses BCC} shows the average surface stress as a function of temperature. Unlike for FCC metals, a general decreasing trend with temperature is absent here and the affect of temperature is much less pronounced, especially for W and V. The (001) surface shows an increase in surface stress, while the (011) and (111) surfaces exhibit a decreasing surface stress with increasing temperature. Also, the (001) surface has a significantly higher average surface stress compared to the other surfaces of the three materials. The general trend is $\tau^{\text{avg}}_{(001)}>\tau^{\text{avg}}_{(111)}>\tau^{\text{avg}}_{(011)}$ for Fe and $\tau^{\text{avg}}_{(001)}>\tau^{\text{avg}}_{(011)}>\tau^{\text{avg}}_{(111)}$ for W and V at all temperatures studied. Fig.~\ref{fig: Polar plot BCC} shows the polar compliance plots of the surface elastic constants of Fe, W, and V (again showing only quarter or half polar plots by exploiting symmetries). Let us point out some specifics for each surface orientation (for all data, see Appendix~\ref{sec:BCCData}).

The (001) BCC surface has 4mm symmetry, analogous to the FCC case. Our results and those by \citet{grochola2002simulation} predict the surface stress for this face to be significantly higher than those of the other faces. (Note that this is in contrast to the ab-initio data by \citet{schonecker2015thermal}, who reported $\tau^{\text{avg}}_{(011)}>\tau^{\text{avg}}_{(001)}$.) The (011) surface of BCC metals belongs to the 2mm symmetry group. It is interesting to note that the Fe (011) surface shows $\tau_{22}<0$, leading to a compressive surface stress at higher temperature, while the W (011) surface shows $\tau_{22}<0$ at all temperatures (see Table~\ref{BCC 011 table}). Also, $\tau_{11}>\tau_{22}$ holds in general. This may seem counter-intuitive due to the atomic arrangement: every atom has two nearest neighbors on either side of the $y$-$z$-plane under an angle of $\cos^{-1}(1/\sqrt{3}) \approx  55^\circ$ with respect to the $x$-axis, so that the stretch originating from these nearest neighbors have more effect along the $y$-axis. However, two more nearest neighbors lie below the surface in the $x$-$z$-plane (hence contributing to forces only along the $x$-direction) and two second-nearest neighbors on the surface along the $x$-axis, both of which lead to a stronger tensile stress along the $x$-direction. Finally, the (111) surface has 3m symmetry and is isotropic but lacks the closed-packed structure of the (111) FCC surface. Isotropy is verified via $2C_{1212}=C_{1111}-C_{1122}$, which holds up to three significant digits (see Table~\ref{BCC 111 table}).

Like for FCC metals, common trends of surface elastic constants with temperature are difficult to identify for BCC metals. 

\section{Conclusion}
\label{conclusion}

We have presented a computationally efficient methodology to obtain surface free energies, surface stresses, and surface elastic constants at finite temperature from a statistical-mechanics-based Gaussian phase packet approach for quasistatic atomistic calculations and asymptotically converged large-thickness values of the surface free energy. This approach allows us to escape the cumbersome path integration or thermodynamic integration procedures used to obtain free energies from MD. We used this setup to obtain finite-temperature surface properties of differently oriented surfaces of representative FCC and BCC metals. Validation has been achieved by comparison to data from MD thermodynamic integration for selected materials, which shows overall excellent agreement (using the same interatomic potentials in LAMMPS). We also reported good agreement with existing literature data, where available. Even though the actual values of surface energy do not match closely with finite-temperature ab-initio calculations (which is an artifact of the limitations of EAM potentials), trends with temperature are correctly reproduced. We point out that, although results have been obtained with EAM potentials, the approach is sufficiently general to apply to more sophisticated potentials -- such as, e.g., potentials of the modified embedded atom method (MEAM). 

The surface property trends observed in this work are summarized as follows. All surface free energies decrease with increasing temperature, which is consistent with previous ab-initio \cite{schonecker2015thermal,wang2019effects} and MC/MD \cite{grochola2002simulation,grochola2004exact,frolov2009temperature,xing2020temperature} calculations. Surface stresses also decrease with increasing temperature. There are, however, some anomalous trends. These include the Al (001) and (111) surfaces and the BCC (001) surfaces of Fe, W, and V. These trends match well with previous MC/MD calculations \cite{frolov2009temperature,grochola2002simulation}, although not well with the ab-initio results of \citet{schonecker2015thermal}. Finally, we have presented the surface elastic constants for each BCC/FCC metal and surface orientation studied, which do not exhibit any common trends with temperature. Different surfaces show interesting variations of the directional Young's modulus, which we have illustrated by polar plots of the directional compliance. We are not aware of prior studies reporting the surface elastic constants of FCC and BCC metals with varying temperature, which could serve for reference.

In short, the GPP approach can bypass cumbersome and expensive MD calculations by computing equilibrium properties in a quasistatic relaxation step. Various problems related to metallic surfaces can be studied based by the presented GPP and asymptotic surface free energy approach, including the study of reconstructed or disordered surfaces. 
An example of the latter is Fe (111) \cite{grochola2002new}, which is known to show adatom hops and faceting at elevated temperature. Similarly, \citet{freitas2017step} studied the step formation energy on a Cu (111) surface using MD, which could alternatively by treated by our methodology, including the quest for stability criteria. Another interesting extension will be to study the effect of alloying on surface energies and surface stresses \cite{wang2019effects}.

\section*{Acknowledgments}
The support from the European Research Council (ERC) under the European Union’s Horizon 2020 research and innovation program (grant agreement no.~770754) is gratefully acknowledged.

\section*{Declaration of Competing Interest}
The authors declare that they have no known competing financial interests or personal relationships that could have appeared to influence the work reported in this paper.

\section*{Data availability statement}
Research data will be made available to interested scientists upon reasonable request. All surface stress and elastic constant data generated during this study are included in this article.

\appendix
\section*{Appendix}


\section{FCC Data}
\label{sec:FCCData}
\vspace{24pt}
\begin{center}
\begin{tabu}{| m{0.7cm} | m{0.7cm} |  m{1.1cm} | m{1.0cm} | m{1.0cm} | m{0.1cm} | m{0.7cm} | m{0.7cm} |  m{1.1cm} | m{1.1cm} | m{1.1cm} |} \cline{1-5} \cline{7-11} & & Cu & Al & Ni &  &  &  & Cu & Al & Ni \\ 
\cline{1-5} \cline{7-11} \multirow{3}{0.4cm}{$\tau_{11}$} & 0~K & 0.0937 & 0.0539 & 0.1225 & & \multirow{3}{0.7cm}{$C_{1111}$} & 0~K & -0.1361 & -0.1427 & -0.2973 \\ 
\cline{2-5} \cline{8-11}  & 100~K & 0.0920 & 0.0551 & 0.1227 & &  & 100~K & -0.1410 & -0.2584 & -0.2940  \\
\cline{2-5} \cline{8-11}  & 300~K & 0.0887 & 0.0576 & 0.1180 & &  & 300~K & -0.1495 & -0.6253 & -0.2180 \\
\cline{2-5} \cline{8-11}  & 500~K & 0.0855 & 0.0575 & 0.1133 & &  & 500~K & -0.1643 & -0.7693 & -0.2275 \\ 
\cline{1-5} \cline{7-11} \multirow{3}{0.4cm}{$\tau_{22}$} & 0~K & 0.0937 & 0.0539 & 0.1225 & & \multirow{3}{0.7cm}{$C_{2222}$} & 0~K & -0.1361 & -0.1427 & -0.2973 \\ 
\cline{2-5} \cline{8-11}  & 100~K & 0.0920 & 0.0551 & 0.1227 & &  & 100~K & -0.1410 & -0.2584 & -0.2940 \\
\cline{2-5} \cline{8-11}   & 300~K & 0.0887 & 0.0576 & 0.1180 & &  & 300~K & -0.1495 & -0.6253 & -0.2180 \\
\cline{2-5} \cline{8-11}  & 500~K & 0.0855 & 0.0575 & 0.1133 & &  & 500~K & -0.1643 & -0.7693 & -0.2275 \\
\cline{1-5} \cline{7-11} \multirow{3}{0.4cm}{$\tau_{12}$} & 0~K & 0.0 & 0.0 & 0.0 & & \multirow{3}{0.7cm}{$C_{1122}$} & 0~K & 0.1508 & 0.1598 & 0.2845 \\ 
\cline{2-5} \cline{8-11}   & 100~K & 0.0 & 0.0 & 0.0 & &  & 100~K & 0.1583 & 0.1015 & 0.2777 \\ 
\cline{2-5} \cline{8-11}  & 300~K & 0.0 & 0.0 & 0.0 & &  & 300~K & 0.1725 & -0.0421 & 0.3011 \\
\cline{2-5} \cline{8-11}  & 500~K & 0.0 & 0.0 & 0.0 & &  & 500~K & 0.1804 & -0.1115 & 0.2439 \\
\cline{1-5} \cline{7-11} \multirow{3}{0.4cm}{$C_{1212}$} & 0~K & -0.0077 & 0.3287 & 0.0432 & & \multirow{3}{0.7cm}{$C_{2211}$} & 0~K & 0.1508 & 0.1598 & 0.2845 \\
\cline{2-5} \cline{8-11}  & 100~K & 0.0006 & 0.2939 & 0.0309 & &  & 100~K & 0.1583 & 0.1015 & 0.2777 \\
\cline{2-5} \cline{8-11}  & 300~K & 0.0153 & 0.1569 & 0.0550 & &  & 300~K & 0.1725 & -0.0421 & 0.3011 \\
\cline{2-5} \cline{8-11}  & 500~K & 0.0271 & 0.1480 & 0.1027 & &  & 500~K & 0.1804 & -0.1115 & 0.2439 \\ 
\cline{1-5} \cline{7-11}
\end{tabu}
\captionof{table}{Surface stress tensor and elastic modulus tensor components of FCC metals for the (001)  crystal surface. (All values are in $eV/$\AA$^2$.)}
\label{FCC 001 table}
\end{center}
\vspace{24pt}
\begin{center}
  \begin{tabu}{| m{0.7cm} | m{0.7cm} |  m{1.1cm} | m{1.1cm} | m{1.1cm} | m{0.1cm} | m{0.7cm} | m{0.7cm} |  m{1.1cm} | m{1.1cm} | m{1.1cm} |} 
  \cline{1-5} \cline{7-11} & & Cu & Al & Ni &  &  &  & Cu & Al & Ni \\ 
  \cline{1-5} \cline{7-11} \multirow{3}{0.4cm}{$\tau_{11}$} & 0~K & 0.0831 & 0.0721 & 0.0978 & & \multirow{3}{0.7cm}{$C_{1111}$} & 0~K & -0.6549 & -0.9959 & -1.1173 \\ 
  \cline{2-5} \cline{8-11}  & 100~K & 0.0823 & 0.0704 & 0.0975 & &  & 100~K & -0.6747 & -0.9985 & -1.1019  \\ 
  \cline{2-5} \cline{8-11}  & 300~K & 0.0807 & 0.0665 & 0.0933 & &  & 300~K & -0.7090 & -1.1398 & -1.0453  \\ 
  \cline{2-5} \cline{8-11}  & 500~K & 0.0790 & 0.0626 & 0.0891 & &  & 500~K & -0.7368 & -1.1650 & -1.0011 \\
  \cline{1-5} \cline{7-11} \multirow{3}{0.4cm}{$\tau_{22}$} & 0~K & 0.0737 & 0.1028 & 0.0851 & & \multirow{3}{0.7cm}{$C_{2222}$} & 0~K & -0.2698 & -0.0592 & -0.5390 \\
  \cline{2-5} \cline{8-11}  & 100~K & 0.0719 & 0.0984 & 0.0850 & &  & 100~K & -0.2771 & -0.0127 & -0.5605 \\
  \cline{2-5} \cline{8-11}   & 300~K & 0.0680 & 0.0902 & 0.0803 & &  & 300~K & -0.2814 & -0.1572 & -0.5375  \\
  \cline{2-5} \cline{8-11}  & 500~K & 0.0637 & 0.0819 & 0.0761 & &  & 500~K & -0.2884 & -0.1260 & -0.4696 \\
  \cline{1-5} \cline{7-11} \multirow{3}{0.4cm}{$\tau_{12}$} & 0~K & 0.0 & 0.0 & 0.0 & & \multirow{3}{0.7cm}{$C_{1122}$} & 0~K & -0.4151 & -0.3599 & -0.7823 \\
  \cline{2-5} \cline{8-11}   & 100~K & 0.0 & 0.0 & 0.0 & &  & 100~K & -0.4298 & -0.3115 & -0.7733  \\ 
  \cline{2-5} \cline{8-11}  & 300~K & 0.0 & 0.0 & 0.0 & &  & 300~K & -0.4503 & -0.3378 & -0.7933  \\
  \cline{2-5} \cline{8-11}  & 500~K & 0.0 & 0.0 & 0.0 & &  & 500~K & -0.4640 & -0.3194 & -0.7696 \\
  \cline{1-5} \cline{7-11} \multirow{3}{0.4cm}{$C_{1212}$} & 0~K & -0.2134 & -0.0295 & -0.5005 & & \multirow{3}{0.7cm}{$C_{2211}$} & 0~K & -0.4057 & -0.3905 & -0.7697 \\ 
  \cline{2-5} \cline{8-11}  & 100~K & -0.2187 & -0.0133 & -0.5057 & &  & 100~K & -0.4193 & -0.3395 & -0.7608 \\
  \cline{2-5} \cline{8-11}  & 300~K & -0.2293 & -0.0551 & -0.5099 & &  & 300~K & -0.4376 & -0.3615 & -0.7802 \\
  \cline{2-5} \cline{8-11}  & 500~K & -0.2384 & -0.0520 & -0.4760 & &  & 500~K & -0.4487 & -0.3386 & -0.7566 \\
  \cline{1-5} \cline{7-11}
\end{tabu}
\captionof{table}{Surface stress tensor and elastic modulus tensor components of FCC metals for the (011)  crystal surface. (All values are in $eV/$\AA$^2$.)}
\label{FCC 011 table}
\end{center}

\begin{center}
\begin{tabu}{| m{0.7cm} | m{0.7cm} |  m{1.1cm} | m{1.1cm} | m{1.1cm} | m{0.1cm} | m{0.7cm} | m{0.7cm} |  m{1.1cm} | m{1.1cm} | m{1.1cm} |}
\cline{1-5} \cline{7-11} & & Cu & Al & Ni &  &  &  & Cu & Al & Ni\\
\cline{1-5} \cline{7-11} \multirow{3}{0.4cm}{$\tau_{11}$} & 0~K & 0.0702 & 0.0766 & 0.0737 & & \multirow{3}{0.7cm}{$C_{1111}$} & 0~K & -4.1603 & 0.3682 & -3.9303 \\ 
\cline{2-5} \cline{8-11}  & 100~K & 0.0684 & 0.0770 & 0.0742 & &  & 100~K & -4.1486 & 0.4359 & -3.8635  \\
\cline{2-5} \cline{8-11}  & 300~K & 0.0645 & 0.0776 & 0.0703 & &  & 300~K & -4.0197 & 0.4187 & -3.5273 \\
\cline{2-5} \cline{8-11}  & 500~K & 0.0606 & 0.0782 & 0.0673 & &  & 500~K & -3.7640 & 0.4253 & -2.5584 \\
\cline{1-5} \cline{7-11} \multirow{3}{0.4cm}{$\tau_{22}$} & 0~K & 0.0702 & 0.0766 & 0.0737 & & \multirow{3}{0.7cm}{$C_{2222}$} & 0~K & -4.1603 & 0.3682 & -3.9303 \\
\cline{2-5} \cline{8-11}  & 100~K & 0.0684 & 0.0770 & 0.0742 & &  & 100~K & -4.1486 & 0.4359 & -3.8635 \\
\cline{2-5} \cline{8-11}  & 300~K & 0.0645 & 0.0776 & 0.0703 & &  & 300~K & -4.0197 & 0.4187 & -3.5273 \\
\cline{2-5} \cline{8-11}  & 500~K & 0.0606 & 0.0782 & 0.0673 & &  & 500~K & -3.7640 & 0.4253 & -2.5584 \\
\cline{1-5} \cline{7-11} \multirow{3}{0.4cm}{$\tau_{12}$} & 0~K & 0.0 & 0.0 & 0.0 & & \multirow{3}{0.7cm}{$C_{1122}$} & 0~K & 4.1298 & 0.4145 & 3.8783 \\
\cline{2-5} \cline{8-11}   & 100~K & 0.0 & 0.0 & 0.0 & &  & 100~K & 4.0868 & 0.3999 & 3.8008  \\
\cline{2-5} \cline{8-11}  & 300~K & 0.0 & 0.0 & 0.0 & &  & 300~K & 3.9253 & 0.3722 & 3.5051 \\
\cline{2-5} \cline{8-11}  & 500~K & 0.0 & 0.0 & 0.0 & &  & 500~K & 3.6418 & 0.3826 & 2.6173 \\
\cline{1-5} \cline{7-11} \multirow{3}{0.4cm}{$C_{1212}$} & 0~K & -4.1406 & -0.0183 & -3.8995 & & \multirow{3}{0.7cm}{$C_{2211}$} & 0~K & 4.1298 & 0.4145 & 3.8783 \\
\cline{2-5} \cline{8-11}  & 100~K & -4.1135 & 0.0227 & -3.8275 & &  & 100~K & 4.0868 & 0.3999 & 3.8008 \\ 
\cline{2-5} \cline{8-11}  & 300~K & -3.9685 & 0.0281 & -3.5114 & &  & 300~K & 3.9253 & 0.3722 & 3.5051 \\
\cline{2-5} \cline{8-11}  & 500~K & -3.6991 & 0.0262 & -2.5834 & &  & 500~K & 3.6418 & 0.3826 & 2.6173 \\
\cline{1-5} \cline{7-11}
\end{tabu}
\captionof{table}{Surface stress tensor and elastic modulus tensor components of FCC metals for the (111) crystal surface. (All values are in $eV/$\AA$^2$.)}
\label{FCC 111 table}
\end{center}

\section{BCC Data}
\label{sec:BCCData}
\vspace{24pt}
\begin{center}
\begin{tabu}{| m{0.7cm} | m{0.7cm} |  m{1.1cm} | m{1.1cm} | m{1.1cm} | m{0.1cm} | m{0.7cm} | m{0.7cm} |  m{1.1cm} | m{1.1cm} | m{1.1cm} |} \cline{1-5} \cline{7-11} & & Fe & W & V &  &  &  & Fe & W & V \\ 
\cline{1-5} \cline{7-11} \multirow{3}{0.4cm}{$\tau_{11}$} & 0~K & 0.1261 & 0.2118 & 0.1349 & & \multirow{3}{0.7cm}{$C_{1111}$} & 0~K & -0.8617 & -0.9394 & -0.5857 \\ 
\cline{2-5} \cline{8-11}  & 100~K & 0.1273 & 0.2160 & 0.1349 & &  & 100~K & -0.8776 & -0.9560 & -0.5979 \\
\cline{2-5} \cline{8-11}  & 300~K & 0.1276 & 0.2231 & 0.1345 &  &  & 300~K & -0.7998 & -0.9425 & -0.5814 \\
\cline{2-5} \cline{8-11}  & 500~K & 0.1260 & 0.2257 & 0.1328 & &  & 500~K & -0.7349 & -0.9986 & -0.5661 \\ 
\cline{1-5} \cline{7-11} \multirow{3}{0.4cm}{$\tau_{22}$} & 0~K & 0.1261 & 0.2118 & 0.1349 & & \multirow{3}{0.7cm}{$C_{2222}$} & 0~K & -0.8617 & -0.9394 & -0.5857 \\ 
\cline{2-5} \cline{8-11}  & 100~K & 0.1273 & 0.2160 & 0.1349 & &  & 100~K & -0.8776 & -0.9560 & -0.5979 \\
\cline{2-5} \cline{8-11}  & 300~K & 0.1276 & 0.2231 & 0.1345 &  &  & 300~K & -0.7998 & -0.9425 & -0.5814 \\
\cline{2-5} \cline{8-11}  & 500~K & 0.1260 & 0.2257 & 0.1328 & &  & 500~K & -0.7349 & -0.9986 & -0.5661 \\ 
\cline{1-5} \cline{7-11} \multirow{3}{0.4cm}{$\tau_{12}$} & 0~K & 0.0 & 0.0 & 0.0 & & \multirow{3}{0.7cm}{$C_{1122}$} & 0~K & -0.8521 & -1.1688 & -0.5484 \\ 
\cline{2-5} \cline{8-11}   & 100~K & 0.0 & 0.0 & 0.0 & &  & 100~K & -0.8597 & -1.1973 & -0.5397 \\ 
\cline{2-5} \cline{8-11}  & 300~K & 0.0 & 0.0 & 0.0 & &  & 300~K & -0.8400 & -1.1763 & -0.5462 \\
\cline{2-5} \cline{8-11}  & 500~K & 0.0 & 0.0 & 0.0 & &  & 500~K & -0.8361 & -1.1628 & -0.5395 \\
\cline{1-5} \cline{7-11} \multirow{3}{0.4cm}{$C_{1212}$} & 0~K & -0.8293 & -0.3925 & -0.0114 & & \multirow{3}{0.7cm}{$C_{2211}$} & 0~K & -0.8521 & -1.1688 & -0.5484 \\
\cline{2-5} \cline{8-11}  & 100~K & -0.8406 & -0.4128 & -0.0375 & &  & 100~K & -0.8597 & -1.1973 & -0.5397 \\
\cline{2-5} \cline{8-11}  & 300~K & -0.8605 & -0.4266 & -0.0171 & &  & 300~K & -0.8400 & -1.1763 & -0.5462 \\
\cline{2-5} \cline{8-11}  & 500~K & -0.8817 & -0.4563 & -0.0215 & &  &  500~K & -0.8361 & -1.1628 & -0.5395 \\ 
\cline{1-5} \cline{7-11}
\end{tabu}
\captionof{table}{Surface stress tensor and elastic modulus tensor components of BCC metals for the (001)  crystal surface. (All values are in $eV/$\AA$^2$.)}
\label{BCC 001 table}
\end{center}
\begin{center}
\begin{tabu}{| m{0.7cm} | m{0.7cm} |  m{1.1cm} | m{1.1cm} | m{1.1cm} | m{0.1cm} | m{0.7cm} | m{0.7cm} |  m{1.1cm} | m{1.1cm} | m{1.1cm} |} \cline{1-5} \cline{7-11} & & Fe & W & V &  &  &  & Fe & W & V  \\ 
\cline{1-5} \cline{7-11} \multirow{3}{0.4cm}{$\tau_{11}$} & 0~K & 0.0608 & 0.1554 & 0.1138 & & \multirow{3}{0.7cm}{$C_{1111}$} & 0~K & 0.0450 & -0.1107 & -0.1196 \\ 
\cline{2-5} \cline{8-11}  & 100~K & 0.0591 & 0.1567 & 0.1114 & &  & 100~K & 0.0414 & -0.0978 & -0.1281 \\
\cline{2-5} \cline{8-11}  & 300~K & 0.0556 & 0.1592 & 0.1099 &  &  & 300~K & 0.0256 & -0.0816 & -0.1052 \\
\cline{2-5} \cline{8-11}  & 500~K & 0.0521 & 0.1596 & 0.1075 & &  & 500~K & 0.0314 & -0.1407 & -0.1025 \\ 
\cline{1-5} \cline{7-11} \multirow{3}{0.4cm}{$\tau_{22}$} & 0~K & 0.0083 & -0.0036 & 0.0332 & & \multirow{3}{0.7cm}{$C_{2222}$} &  0~K & -0.3779 & -0.6250 & -0.2317 \\
\cline{2-5} \cline{8-11}  & 100~K & 0.0021 & -0.0073 & 0.0300 & & & 100~K & -0.3430 & -0.6117 & -0.2077 \\
\cline{2-5} \cline{8-11}  & 300~K & -0.0080 & -0.0146 & 0.0282 & &  & 300~K & -0.2924 & -0.6053 & -0.1619 \\
\cline{2-5} \cline{8-11}  & 500~K & -0.0147 & -0.0200 & 0.0265 & &  & 500~K & -0.2122 & -0.6465 & -0.1458 \\ 
\cline{1-5} \cline{7-11} \multirow{3}{0.4cm}{$\tau_{12}$} & 0~K & 0.0 & 0.0 & 0.0 & & \multirow{3}{0.7cm}{$C_{1122}$} & 0~K & 0.3496 & 0.4289 & 0.0107 \\ 
\cline{2-5} \cline{8-11}   & 100~K & 0.0 & 0.0 & 0.0 & &  & 100~K & 0.3744 & 0.4607 & 0.0369 \\ 
\cline{2-5} \cline{8-11}  & 300~K & 0.0 & 0.0 & 0.0 & &  & 300~K & 0.4007 & 0.4590 & 0.0606 \\
\cline{2-5} \cline{8-11}  & 500~K & 0.0 & 0.0 & 0.0 & &  & 500~K & 0.4426 & 0.4378 & 0.0584 \\
\cline{1-5} \cline{7-11} \multirow{3}{0.4cm}{$C_{1212}$} & 0~K & -0.0422 & 0.1186 & -0.1184 & & \multirow{3}{0.7cm}{$C_{2211}$} & 0~K & 0.4022 & 0.5879 & 0.0913 \\
\cline{2-5} \cline{8-11}  & 100~K & 0.0102 & 0.1465 & -0.0953 & &  & 100~K & 0.4314 & 0.6248 & 0.1183 \\
\cline{2-5} \cline{8-11}  & 300~K & 0.0922 & 0.1602 & -0.0777 & &  & 300~K & 0.4643 & 0.6328 & 0.1423 \\
\cline{2-5} \cline{8-11}  & 500~K & 0.1496 & 0.1662 & -0.0775 & &  &  500~K & 0.5094 & 0.6174 & 0.1394 \\ 
\cline{1-5} \cline{7-11}
\end{tabu}
\captionof{table}{Surface stress tensor and elastic modulus tensor components of BCC metals for the (011)  crystal surface. (All values are in $eV/$\AA$^2$.)}
\label{BCC 011 table}
\end{center}
\begin{center}
\begin{tabu}{| m{0.7cm} | m{0.7cm} |  m{1.1cm} | m{1.1cm} |  m{1.1cm} | m{0.1cm} | m{0.7cm} | m{0.7cm} |  m{1.1cm} | m{1.1cm} |  m{1.1cm} |} \cline{1-5} \cline{7-11} & & Fe & W & V & &  &  & Fe & W & V \\ 
\cline{1-5} \cline{7-11} \multirow{3}{0.4cm}{$\tau_{11}$} & 0~K & 0.0451 & 0.0241 & 0.0285 & & \multirow{3}{0.7cm}{$C_{1111}$} & 0~K & -4.2446 & -1.4018 & -0.3523 \\ 
\cline{2-5} \cline{8-11}  & 100~K & 0.0423 & 0.0217 & 0.0259 & &  & 100~K & -4.6476 & -1.5115 & -0.3118 \\
\cline{2-5} \cline{8-11}  & 300~K & 0.0380 & 0.0172 & 0.0254 &  &  & 300~K & -4.8656 & -1.5367 & -0.2608 \\
\cline{2-5} \cline{8-11}  & 500~K & 0.0347 & 0.0150 & 0.0248 & &  & 500~K & -4.4122 & -1.6257 & -0.2640 \\ 
\cline{1-5} \cline{7-11} \multirow{3}{0.4cm}{$\tau_{22}$} & 0~K & 0.0451 & 0.0241 & 0.0285 & & \multirow{3}{0.7cm}{$C_{2222}$} &  0~K & -4.2446 & -1.4018 & -0.3523 \\ 
\cline{2-5} \cline{8-11}  & 100~K & 0.0423 & 0.0217 & 0.0259 & &  & 100~K & -4.6476 & -1.5115 & -0.3118 \\
\cline{2-5} \cline{8-11}  & 300~K & 0.0380 & 0.0172 & 0.0254 &  &  & 300~K & -4.8656 & -1.5367 & -0.2608 \\
\cline{2-5} \cline{8-11}  & 500~K & 0.0347 & 0.0150 & 0.0248 & &  & 500~K & -4.4122 & -1.6257 & -0.2640 \\ 
\cline{1-5} \cline{7-11} \multirow{3}{0.4cm}{$\tau_{12}$} & 0~K & 0.0 & 0.0 & 0.0 & & \multirow{3}{0.7cm}{$C_{1122}$} & 0~K & 2.1426 & 0.3512 & 0.1642 \\ 
\cline{2-5} \cline{8-11}   & 100~K & 0.0 & 0.0 & 0.0 & &  & 100~K & 2.5696 & 0.2946 & 0.0707 \\ 
\cline{2-5} \cline{8-11}  & 300~K & 0.0 & 0.0 & 0.0 & &  & 300~K & 2.8431 & 0.2290 & 0.1270 \\
\cline{2-5} \cline{8-11}  & 500~K & 0.0 & 0.0 & 0.0 & &  & 500~K & 2.4307 & 0.1883 & 0.1713 \\
\cline{1-5} \cline{7-11} \multirow{3}{0.4cm}{$C_{1212}$} & 0~K & -3.1903 & -0.8709 & -0.2565 & & \multirow{3}{0.7cm}{$C_{2211}$} & 0~K & 2.1426 & 0.3512 & 0.1642 \\
\cline{2-5} \cline{8-11}  & 100~K & -3.6026 & -0.8998 & -0.1895 & &  & 100~K & 2.5696 & 0.2946 & 0.0707 \\
\cline{2-5} \cline{8-11}  & 300~K & -3.8559 & -0.8823 & -0.1929 & &  & 300~K & 2.8431 & 0.2290 & 0.1270 \\
\cline{2-5} \cline{8-11}  & 500~K & -3.4208 & -0.9072 & -0.2159 & &  &  500~K & 2.4307 & 0.1883 & 0.1713 \\ 
\cline{1-5} \cline{7-11}
\end{tabu}
\captionof{table}{Surface stress tensor and elastic modulus tensor components of BCC metals for the (111)  crystal surface. (All values are in $eV/$\AA$^2$.)}
\label{BCC 111 table}
\end{center}



\clearpage

\bibliographystyle{cas-model2-names}
\bibliography{surface_energy}

\end{document}